\documentstyle[12pt]{article}

\newcommand{\sect}[1]{\section{#1}}
\newcommand{\lbl}[1]{\label{eq:#1}}
\newcommand{\rf}[1]{(\ref{eq:#1})}
\newcommand{\vs}[1]{\rule[- #1 mm]{0mm}{#1 mm}}
\newcommand{\eq}{\vs{2}\begin{equation}}
\newcommand{\en}{\\[2mm]\end{equation}}
\newcommand{\bea}{\begin{eqnarray}}
\newcommand{\ena}{\end{eqnarray}}
\newcommand{\lapprox}{%
\mathrel{%
\setbox0=\hbox{$<$}
\raise0.6ex\copy0\kern-\wd0
\lower0.65ex\hbox{$\sim$}
}}
\newcommand{\gapprox}{%
\mathrel{%
\setbox0=\hbox{$>$}
\raise0.6ex\copy0\kern-\wd0
\lower0.65ex\hbox{$\sim$}
}}

\newcommand{\NP}[1]{Nucl.\ Phys.\ {\bf #1}}

\newcommand{\PL}[1]{Phys.\ Lett.\ {\bf #1}}
\newcommand{\NC}[1]{Nuovo Cimento {\bf #1}}
\newcommand{\AN}[1]{Ann. Phys. {\bf #1}}

\newcommand{\PRev}[1]{Phys.\ Rev.\ {\bf #1}}
\newcommand{\PRL}[1]{Phys.\ Rev.\ Lett.\ {\bf #1}}

\def\app{\alpha_{\pi\pi}}
\def\bpp{\beta_{\pi\pi}}
\def\apik{\alpha_{\pi K}}
\def\bpik{\beta_{\pi K}}
\def\mpi{M_\pi}
\def\fpi{F_\pi}
\def\mk{M_K}

\def\cpt{$\chi{\rm PT}\ $}
\def\arctanh{ {\rm arctanh}}

\begin{document}
\setcounter{page}{0}
\renewcommand{\thefootnote}{\fnsymbol{footnote}}

\rightline{IPNO/TH 94-08}

\indent

\indent

\begin{center}
{\Large {\bf The Reaction $\gamma\gamma\to\pi^0\pi^0$ }}

\indent

{\Large {\bf in Generalized Chiral Perturbation Theory }}

\vskip 3 true cm

{\large M. Knecht, B. Moussallam, J. Stern }

\indent

{\sl Division de Physique Th\'eorique
\footnote{Unit\'e de Recherche des Universit\'es Paris XI et Paris VI
associ\'ee au CNRS.}, Institut de Physique Nucliaire\\
F-91406 Orsay Cedex, France} \\

\indent

\end{center}

\vskip 4 true cm

\centerline{\large {\bf Abstract}}

\indent

The cross section for $\gamma\gamma\to\pi^0\pi^0$ and the pion polarizabilities
are computed, within generalized chiral perturbation theory, in the full one
loop approximation, {\it i.e.} up to and including order $O({p^5})$.
The result depends on the parameter ${\alpha}_{\pi\pi}$ defining the tree
level $\pi - \pi$ scattering amplitude and on an additional
low energy constant. The latter is shown to be related by an exact sum-rule
to the $e^+ e^-$ data. The parameter ${\alpha}_{\pi\pi}$ is related to the
quark mass ratio $r = m_s /{\hat m}$ via the expansion of pseudoscalar meson
masses. The generalized one loop $\gamma\gamma\to\pi^0\pi^0$ amplitude agrees
with the experimental data in the threshold region provided
$r= {m_s}/{\hat m}\lapprox 10$. Higher order corrections are estimated
comparing our calculation with the dispersive approach.

\indent

\indent

\noindent IPNO/TH 94-08 \\
January 1994

\newpage


\renewcommand{\thefootnote}{\arabic{footnote}}
\setcounter{footnote}{0}

\sect{Introduction}

\indent

The reaction $\gamma\gamma\to{\pi}^0{\pi}^0$ at low energies provides a so far
rare example of a probe of the chiral symmetry breaking sector of QCD for which
experimental data already exist \cite{data}. Moreover, these data do not
exhibit any unexpected feature, as they can be reproduced without difficulty
\cite{GRR}-\cite{russes} using the standard methods based on dispersion
relations, unitarity and resonance saturation. The one-loop prediction of the
standard chiral perturbation theory \cite{GL} ($\chi$PT) appeared prior
\cite{BC,DHL} to the publication of the experimental results. It does not
involve any free parameter and it disagrees with the data by several standard
deviations even close to threshold. Recently, the full two-loop calculation
has been completed by Bellucci, Gasser and Sainio \cite{BGS}. It involves
three new $O(p^6 )$ constants (estimated via resonance saturation) and it
agrees with the experimental data within errors. The purpose of this work is
to add one more information to this list: In the framework of
{\it generalized} $\chi$PT (defined in \cite{FSS2,SSF}), the agreement with
experiment near threshold is already reached within the one-loop approximation,
provided the ratio of quark masses $r\equiv {m_s}/{\hat m}$ is considerably
lower than usually expected, typically $r\le 10$.

The important question to ask is: What can $\gamma\gamma\to \pi^0\pi^0$ data
tell us about the expansion of the QCD effective Lagrangian \cite{W,GLlett}?
While each individual term of ${\cal L}^{eff.}$ is uniquely
determined \cite{L} by symmetry properties of QCD, the relative importance of
different terms depends on the actual values of the quark masses and of the
low-energy constants
characterizing the chiral structure of the QCD vacuum. The latter in turn
determines how the expansion of ${\cal L}^{eff.}$ in powers of quark masses and
external momenta should be organized such as to obtain a good convergence rate.
The {\it standard chiral perturbation theory} \cite{GL} is an expansion which
assumes that the quark masses $m_u$, $m_d$ and $m_s$ are small enough not only
with respect to the hadronic scale ${\Lambda}_H \sim 1$ GeV, but also
as compared to the scale $B_0$ of the single flavour quark-antiquark
condensate, $<{\bar q}q>_0
=- B_0 F_0^2$, defined at $m_u = m_d = m_s = 0$. Within QCD, the last
assumption
is hard to justify {\it a priori}, since nothing prevents $B_0$ to be as small
as, say, the pion decay constant $F_0 \sim 90$ MeV, {\it i.e.} much smaller
than
${\Lambda}_H$. To illustrate the importance of the condition $m_q \ll B_0$,
let us mention the fact that the standard relation between pseudoscalar meson
and quark mass ratios \cite{W2} :
\eq
r = {{m_s}\over{\hat m}} = 2\,{{M_K^2}\over{M_{\pi}^2}} - 1 + \cdots = 25.9 +
\cdots\ ,\quad {\hat m}\equiv {1\over 2}(m_u + m_d )\ ,\lbl{r}
\en
receives, at the next order, corrections \cite{GL,L2} (represented by the
ellipses) which are of both types, $O({m_q}/{\Lambda}_H )$ {\it and}
$O({m_q}/B_0 )$. On the other hand, the analysis of the Dashen-Weinstein
sum-rule for the deviations from the Goldberger-Treiman relation suggests that
$r$ might differ considerably from the above value \cite{FSS1}, {\it e.g.}
$r\le
10$. If confirmed\footnote{For a recent determination of the $\pi$N coupling
constant, see Ref. \cite{MKB}}, this result could be interpreted as an
experimental indication of an important $O({m_q}/B_0 )$ contribution to Eq.
\rf{r}.
The {\it generalized chiral perturbation theory} \cite{FSS2,SSF}
reformulates the expansion of the effective Lagrangian without assuming that
$m_q\ll B_0$. It is as systematic and consistent as the standard $\chi$PT. At
all orders in $p/{\Lambda}_H$ and in ${m_q}/{\Lambda}_H$, both expansion
schemes, standard and generalized, sum up the same effective Lagrangian
${\cal L}^{eff.}$ of QCD. However, at each {\it finite} order, the generalized
$\chi$PT takes into account terms which standard $\chi$PT relegates to higher
orders. Already at the leading $O(p^2 )$ order, the generalized $\chi$PT
involves more terms and more parameters: In particular, the quark mass ratio
$r$ is a free parameter, to be determined by
experiment. In the limit $r=r_2\equiv 2{M_K^2}/{M_{\pi}^2} - 1$, one recovers
the standard $\chi$PT as a special case. In practice, the generalized $\chi$PT
is defined by the formal counting rule ${\hat m}$, $m_s$, $B_0\, \sim\,O(p)$
\cite{FSS2,SSF}, replacing the usual rule ${\hat m}$,
$m_s\,\sim\,O({p^2}/{\Lambda}_H )$, $B_0\,\sim\,O({\Lambda}_H )$
characteristic of standard $\chi$PT \cite{GL}.
Hence, for $p\ll{\Lambda}_H$, one can write
\eq
{\cal L}^{eff.}\ = {\cal L}^{(2)} + {\cal L}^{(4)} + {\cal L}^{(6)} +\cdots
                      \ = \ {\tilde{\cal L}}^{(2)} + {\tilde{\cal L}}^{(3)} +
{\tilde{\cal L}}^{(4)} + {\tilde{\cal L}}^{(5)} + {\tilde{\cal L}}^{(6)}
+\cdots\ ,
\en
where the first expansion is the standard one \cite{GL}, whereas the second one
corresponds to the generalized $\chi$PT.

In the next section, we summarize our results for the one-loop generalized
$\chi$PT $\gamma\gamma\to\pi^0\pi^0$ amplitude. It depends on the generalized
tree-level $\pi - \pi$ amplitude \cite{FSS2,SSF} (kaon loops give only a
small contribution) and it receives a constant shift from an order $O(p^5 )$
tree-level contribution. The latter is shown to be calculable, through a
low-energy theorem, from $e^+ e^- \to$ hadrons data (Section 3). As an
indicative estimate of higher order corrections, we submit, in Sect. 4, our
result to a dispersive analysis along the lines of Ref. \cite{DH}. We finally
add a few comments on charged and neutral pion polarizabilities.

We  find it convenient to perform the whole analysis within the
three-light-flavours $\chi$PT. It allows us to keep kaon loops under control
and, mainly, it should help in establishing relationships to other observables
(c.f. Sect. 3) and to similar processes such as $\eta\to\pi^0\gamma\gamma$
\cite{Amet}.

\sect{ $\gamma\gamma\to\pi^0\pi^0$ to generalized one loop order}

\indent

In the present section, we establish the expression at generalized one loop
order for the amplitude of the reaction $\gamma\gamma\to\pi^0\pi^0$.
The modified chiral counting specific to generalized $\chi$PT leads to the
appearance of terms of {\it odd} orders, which do not correspond to an
increase in the number of loops, but to additional corrections in powers of
the quark masses. Thus, the leading order contributions from ${\tilde{\cal
L}}^{(2)}$ receive tree-level corrections from ${\tilde{\cal L}}^{(3)}$, which
come before the one loop effects. Similarly, before going to the two loop
order, one has to take into account contributions at order $O(p^5 )$. The
generating functional (S-matrix) at one loop order in generalized $\chi$PT is
thus given by the following expansion:
\eq
{\cal Z}^{eff.} = {\tilde{\cal Z}}_{tree}^{(2)} + {\tilde{\cal Z}}_{tree}^{(3)}
+ {\tilde{\cal Z}}_{tree}^{(4)} + {\tilde{\cal Z}}_{1loop}^{(4)} +
{\tilde{\cal Z}}_{tree}^{(5)} + {\tilde{\cal Z}}_{1loop}^{(5)} +\cdots\ .
\lbl{Zeff}
\en
Here, ${\tilde{\cal Z}}_{tree}^{(2)} $ corresponds to all the tree diagrams
made from an arbitrary number of vertices from ${\tilde{\cal L}}^{(2)} $; $
{\tilde{\cal Z}}_{tree}^{(3)}$ denotes the same set of diagrams, but where
a {\it single} vertex from ${\tilde{\cal L}}^{(2)}$ is replaced by a
vertex from ${\tilde{\cal L}}^{(3)}$. ${\tilde{\cal Z}}_{tree}^{(4)}$ contains
a single vertex from ${\tilde{\cal L}}^{(4)}$ or {\it two} vertices from
${\tilde{\cal L}}^{(3)}$, with an arbitrary number of additional vertices from
${\tilde{\cal L}}^{(2)}$. Similarly, ${\tilde{\cal Z}}_{tree}^{(5)}$ stands for
all tree diagrams made from any number of vertices from ${\tilde{\cal
L}}^{(2)}$ and either a single vertex from ${\tilde{\cal L}}^{(5)}$ or one
vertex from ${\tilde{\cal L}}^{(3)}$ {\it and} one vertex from ${\tilde{\cal
L}}^{(4)}$. Finally, ${\tilde{\cal Z}}_{1loop}^{(4)}$ stands for all one loop
diagrams made with vertices from ${\tilde{\cal L}}^{(2)}$ only, while
${\tilde{\cal Z}}_{1loop}^{(5)}$ denotes the same diagrams, but with one of
those vertices replaced by a vertex from ${\tilde{\cal L}}^{(3)}$.

The general analysis of ${\cal Z}^{eff.}$ up to and including the order
$O(p^5 )$ (Eq. \rf{Zeff}) and of the various low-energy constants involved is
beyond the scope of the present work, and will be given elsewhere
\cite{KS}\footnote{For a discussion of the leading order, see Ref.
\cite{FSS2}} . Here, we shall merely consider those terms which contribute to
the matrix element
\eq
<\,{\pi}^{0}(p_1 ){\pi}^{0}(p_2 )\,out\vert\,{\gamma}(k_1 ,\,{\epsilon}_1
){\gamma}(k_2 ,\,{\epsilon}_2 )\,in> = ie^2 (2\pi )^4{\delta}^4 (P_f - P_i )
{\epsilon}_1^{\mu}{\epsilon}_2^{\nu} M_{\mu\nu}\ ,
\en
with\footnote{The amplitude denoted by $H(s,\,t,\,u)$ is called
$A(s,\,t,\,u)$ by the authors of Ref. \cite{BGS}. We have adopted a different
notation in order to avoid confusion with the standard notation
$A(s\,\vert\,t,\,u)$ for the $\pi - \pi$ scattering amplitude.}
\eq
M_{\mu\nu} = H(s,\,t,\,u) ({s\over 2}{\eta}_{\mu\nu} - k_{1\nu}k_{2\mu}) +
\cdots\ .
\en
(The ellipses stand for the other Lorentz tensors in the general decomposition
of $M_{\mu\nu}$, whose form factors receive contributions, for on-shell
photons, only from order $O(p^6 )$ onward.) In this case, the terms
${\tilde{\cal Z}}_{tree}^{(2)}$, ${\tilde{\cal Z}}_{tree}^{(3)}$ and
${\tilde{\cal Z}}_{tree}^{(4)}$ in Eq.
\rf{Zeff} do not contribute. The remaining three terms contribute to the
amplitude $H(s,\,t,\,u)$, the loop diagrams ${\tilde{\cal Z}}_{1loop}^{(4)}$
and
${\tilde{\cal Z}}_{1loop}^{(5)}$ and the tree diagrams of order $O(p^5 )$ are
all separately finite. Topologically, the loop diagrams are the same as in the
standard case \cite{BC}, \cite{DHL}, but the vertices are to be read off from
${\tilde{\cal L}}^{(2)}$ and from ${\tilde{\cal L}}^{(3)}$, where
\bea
{\tilde{\cal L}}^{(2)} &=& {F_0^2\over 4}\bigg\{
                   \langle D^{\mu}U^+ D_{\mu}U\rangle +
        2B_0 \langle {\cal M}_q ( U + U^{+})\rangle\nonumber\\
        &+& A_0 \langle {({\cal M}_q U)}^2 + {({\cal M}_q U^{+})}^2 \rangle
\lbl{l2}\\
        &+& Z_0^S{\langle {\cal M}_q U + {\cal M}_q U^{+}\rangle}^2 + \cdots
\bigg\} \nonumber \ ,
\ena
and
\bea
{\tilde{\cal L}}^{(3)} &=& {F_0^2\over 4}\bigg\{
         {\xi} \langle D^{\mu}U^+ D_{\mu}U ({\cal M}_q U + U^+ {\cal
           M}_q )\rangle\nonumber\\
      &+& {\tilde\xi} \langle D^{\mu}U^+ D_{\mu}U\rangle\langle {\cal M}_q (U^+
+ U)\rangle\nonumber\\
      &+& {\rho_1} \langle {({\cal M}_q U)}^3 + {({\cal M}_q U^{+})}^3 \rangle
\, +\, {\rho_2} \langle {\cal M}_{q}^3 ( U + U^+ ) \rangle\lbl{l3}\\
      &+& {\rho_4} \langle {({\cal M}_q U)}^2 + {({\cal M}_q U^{+})}^2\rangle
\langle {\cal M}_q ( U + U^+ ) \rangle \nonumber\\
&+& {\rho_5} \langle {\cal M}_q^2 \rangle \langle {\cal M}_q ( U + U^+ )
\rangle\, +\, {\rho_7} {\langle {\cal M}_q ( U + U^+ ) \rangle}^3 + \cdots
\bigg\} \nonumber\ ,
\ena
with at most one vertex per diagram coming from ${\tilde{\cal L}}^{(3)}$.
The notations are as given in Ref.\cite{GL}, ${\cal M}_q =$ diag(${\hat m}$,
${\hat m}$, $m_s$) denotes the quark mass matrix and the ellipses stand for
those terms which do not contribute to $H(s,\,t,\,u)$. The tree graph
contributions of  order $O(p^5 )$ are described by\footnote{In standard
$\chi$PT, these terms would be part of the $O(p^6 )$ contributions \cite{BGS}.}
\bea
{\tilde{\cal L}}^{(5)} &=& c_1 \langle F_L^{\mu\nu}{\cal M}_q F_{R\mu\nu}U +
F_R^{\mu\nu} {\cal M}_q F_{L\mu\nu}U^+ \rangle\nonumber\\
             &+& c_2 \langle F_L^{\mu\nu} U^+ F_{R\mu\nu} U {\cal M}_q U +
                             F_R^{\mu\nu} U F_{L\mu\nu} U^+ {\cal M}_q U^+
\rangle\nonumber\\
             &+& c_3 \langle F_R^{\mu\nu}F_{R\mu\nu}( U{\cal M}_q + {\cal M}_q
U^+ ) + F_L^{\mu\nu}F_{L\mu\nu}( U^+ {\cal M}_q + {\cal M}_q U )
\rangle\lbl{l5}\\
             &+& c_4 \langle F_R^{\mu\nu}F_{R\mu\nu} + F_L^{\mu\nu}F_{L\mu\nu}
\rangle \langle {\cal M}_q ( U + U^+ ) \rangle + ... \nonumber\ .
\ena
The final result will only
depend on $F_{\pi}$, $M_{\pi}$, and on a few other independent combinations of
${\hat m}$, of $r$, and of the various low-energy constants occurring in
\rf{l2}, \rf{l3} and \rf{l5}. The relevant vertices are given in Fig. 1. Up to
the order $O(p^5 )$, the amplitude $H(s,\,t,\,u)$ depends only on $s$,
$H(s,\,t,\,u) = H(s) + O(p^6 )$, and takes the following form
\eq
H(s) = A^{\pi -loop}(s) + A^{K-loop}(s) -
{4c\over{F_{\pi}^2}}\ .\lbl{amp}
\en
The pion loop contribution $A^{\pi -loop}$ may be written as
\eq
A^{\pi -loop}(s) = {4\over s}\,\big[\, {\beta}_{\pi\pi}\,{{s - {4\over
                      3}M_{\pi}^2}\over{F_{\pi}^2}} +
  {\alpha}_{\pi\pi}\,{{M_{\pi}^2}\over{3F_{\pi}^2}}\,\big]\,
{\bar G}({s\over{M_{\pi}^2}})\ ,
\en
where ${\bar G}$ denotes the following loop function,
\eq
-16{\pi}^2 {\bar G}(z) = \left\{ \begin{array}{ll}
1 + {1\over z}\big(\ln{{1-\sigma}\over{1+\sigma}} + i\pi{\big)}^2 &,\,\mbox{if
$
4\le z$}\\
1 - {4\over z}{\mbox{arctg}}^2\big({z\over{4-z}}{\big)}^{1\over 2}&,\,\mbox{if
$
0\le z\le 4$}\\
1 + {1\over z}{\ln}^2{{\sigma - 1}\over{\sigma + 1}}              &,\,\mbox{if
$
z\le 0$}\end{array} \right.\ ,\ {\sigma}(z)\equiv {\sqrt{1 - 4/z}}\ .
\en
Here ${\alpha}_{\pi\pi}$ and ${\beta}_{\pi\pi}$ are two combinations of the
low-energy constants of ${\tilde{\cal L}}^{(2)} + {\tilde{\cal L}}^{(3)}$ which
parametrize the $O(p^3 )$ tree-level $\pi - \pi$ scattering amplitude
\cite{FSS2}
\eq
A(s\,\vert\,t,\,u) = {\beta}_{\pi\pi}\,{{s - {4\over
3}M_{\pi}^2}\over{F_{\pi}^2}} +
{\alpha}_{\pi\pi}\,{{M_{\pi}^2}\over{3F_{\pi}^2}} + O(p^4 )\ .
\en
The kaon loop contribution has a similar structure,
\eq
A^{K-loop}(s) = {4\over s}\,\bigg[\,{{{\beta}_{\pi K}}\over{4F_{\pi}^2}
}\,( s - {2\over 3}M_{\pi}^2 - {2\over 3}M_K^2 ) + {1\over{6F_{\pi}^2}}\big[
(M_K - M_{\pi})^2 + 2M_{\pi}M_K{\alpha_{\pi K}}\big]\,\bigg]\,{\bar
G}({s\over{M_K^2}})\ ,
\en
in terms of two other combinations of the low-energy constants, ${\alpha}_{\pi
K}$ and ${\beta}_{\pi K}$, which
parametrize the $O(p^3 )$ tree-level $K - \pi$ amplitude $A^{+}(s,\,t,\,u)$
\footnote{In terms of the $K - \pi$ isospin amplitudes, $A^+ =
{2\over 3}A^{3/2} + {1\over 3}A^{1/2}$.} \cite{KSSF}, symmetric in $s$ and $u$:
\eq
A^{+}(s,\,t,\,u) = {{{\beta}_{\pi K}}\over{4F_{\pi}^2}}\,( t - {2\over
3}M_{\pi}^2 - {2\over 3}M_K^2 ) + {1\over{6F_{\pi}^2}}\big[ (M_K - M_{\pi})^2 +
2M_{\pi}M_K{\alpha_{\pi K}}\big] + O(p^4 )\ .
\en
It is useful to express ${\alpha}_{\pi\pi}$ and ${\alpha}_{\pi K}$ as
\bea
{\alpha}_{\pi\pi}&=& {\alpha}_{\pi\pi}^{lead} +
{\delta}{\alpha}_{\pi\pi}\nonumber\\
{\alpha}_{\pi K}&=& {\alpha}_{\pi K}^{lead} + {\delta}{\alpha}_{\pi K}\ ,
\ena
where ${\alpha}_{\pi\pi}^{lead}$ and ${\alpha}_{\pi K}^{lead}$ arise from
${\tilde{\cal L}}^{(2)}$, whereas ${\delta}{\alpha}_{\pi\pi}$ and
${\delta}{\alpha}_{\pi K}$ stem from ${\tilde{\cal L}}^{(3)}$.
The explicit expressions for ${\delta}{\alpha}_{\pi\pi}$ and for
${\delta}{\alpha}_{\pi K}$ in terms of the low-energy constants appearing in
${\tilde{\cal L}}^{(2)}$ and in
${\tilde{\cal L}}^{(3)}$ will not be used in the sequel and can be obtained
from the formulae collected in the
Appendix. Similarly, one finds:
\eq
{\beta}_{\pi\pi} = 1 + 2\xi{\hat m} + 4{\tilde\xi}{\hat m}\ ,\ \quad
{\beta}_{\pi K} = 1 + 2\xi{\hat m} + 4{\tilde\xi}{\hat m}(3+r)\ .\lbl{beta}
\en
(In both cases, the leading order values are ${\beta}_{\pi\pi}^{lead} =
{\beta}_{\pi K}^{lead} = 1$.)
Finally, we come to the tree contribution from ${\tilde{\cal L}}^{(5)}$ which
gives a constant shift to $H(s)$, see \rf{amp}, with
\eq
c = -{10\over 9}( c_1 + c_2 + 2c_3 ){\hat m} - {16\over 3}{c_4}{\hat m}
\approx -{10\over 9}(c_1 + c_2 + 2c_3 ){\hat m}\ ,\lbl{cst}
\en
where the second, approximate, expression is obtained by invoking the Zweig
rule. This constant shift in $H(s)$ appears (along with other contributions)
only at order $O(p^6 )$ in the standard approach and $c$ is in fact related to
the constant $d_3$ of Ref. \cite{BGS}( see the discussion at the end of
Section 3). However, since the loop contributions to $H(s)$ are finite by
themselves, $c$ is not renormalized at the order we are considering
\footnote{The fact that $c$ is finite in our case, whereas the constant $d_3$
of \cite{BGS} is renormalized does not bear any contradiction, and will be
explained later (Sect. 3).}. The value of $c$ depends on $r$ and will be
discussed below.

First, let us consider the amplitude $H(s)$ when
contributions up to order $O(p^4 )$ alone are taken into account. In that
case, the constants $c$, ${\delta}{\alpha}_{\pi\pi}$ and ${\delta}{\alpha}_{\pi
K}$ drop out, and only the leading, $O(p^2 )$, expressions
for the $\pi - \pi$ and $K - \pi$ amplitudes are involved. These essentially
depend on
$F_{\pi}$, $M_{\pi}$ and $r$, since \cite{FSS2}, \cite{SSF}, \cite{KSSF},
\bea
{\alpha}_{\pi\pi}^{lead} = 1 + 6\,{{r_2 - r}\over{r^2 - 1}}\, (1 + 2\zeta )\ &,
&\
{\beta}_{\pi\pi}^{lead} = 1\  ,\quad r_2\equiv 2\,{{M_K^2}\over{M_{\pi}^2}} - 1
\sim 25.9\ ,
\nonumber\\
{\alpha}_{\pi K}^{lead} = 1 + {{r + 1}\over{r_1 +
1}}\,({\alpha}_{\pi\pi}^{lead} - 1)\ &,&\ {\beta}_{\pi K}^{lead} = 1\
,\quad r_1\equiv\,2{{M_K}\over{M_{\pi}}} - 1\sim 6.3\ ,\lbl{lead}
\ena
and since the Zweig rule violating parameter $\zeta \equiv {Z_0^S}/A_0$ is
expected to be small. In fact, vacuum stability arguments impose the following
bounds at leading order \cite{SSF}:
\eq
r\ge r_1 \ ,
\en
and
\eq
0\le \zeta \le {1\over 2}{{r - r_1}\over{r_2 - r}}\cdot{{r + r_1 + 2}\over{r +
2}}\ .
\en
For $r = r_2$, one recovers the standard values of Weinberg \cite{W2}
${\alpha}_{\pi\pi}^{lead} = {\alpha}_{\pi K}^{lead} = 1$. For the critical
value $r = r_1$, one obtains ${\alpha}_{\pi\pi}^{lead} = {\alpha}_{\pi
K}^{lead} = 4$ \cite{FSS2}, \cite{KSSF}, and for an intermediate value, say
$r\sim$10, $2\le{\alpha}_{\pi\pi}^{lead}\le 2.3$.
Thus, as $r$ decreases from its standard value $r_2$ towards its critical
value $r_1$, ${\alpha}_{\pi\pi}^{lead}$ varies by a factor four. This results
in an increase of $\vert H(s)\vert$ at order $O(p^4 )$ in the  threshold
region of about 25$\%$ when $r$ ranges through the same values. The
corresponding variation of the cross section is shown in Fig. 2.

At full one loop order, i.e. taking into account the order $O(p^5 )$
corrections, the amplitude $H(s)$ no longer depends on $r$ alone. It also
involves the low-energy constants of ${\tilde{\cal L}}^{(3)}$ and the
combination \rf{cst} of low-energy constants from ${\tilde{\cal L}}^{(5)}$. We
first notice that the constant $\xi$ involved in the expressions \rf{beta} for
${\beta}_{\pi\pi}$ and ${\beta}_{\pi K}$ is also responsible for the splitting
between $F_{\pi}$ and $F_K$ at order $O(p^3 )$ \cite{FSS2} (in the standard
case, this splitting occurs only at order $O(p^4 )$):
\eq
F_{\pi}^2 = F_0^2\,\big[ 1 + 2{\hat m}\xi + 2{\hat m}{\tilde \xi} (2+r) +
\cdots\big]\ ,
\en
\eq
F_K^2 = F_0^2\,\big[ 1 + {\hat m}\xi (1+r) + 2{\hat m}{\tilde \xi} (2+r) +
\cdots\big]\ .
\en
Therefore,
\eq
{\hat m}\xi = {1\over{r-1}}\bigg({{F_K^2}\over{F_{\pi}^2}} - 1\bigg) + \cdots\
,\lbl{mxi}
\en
which allows us to express ${\beta}_{\pi\pi}$ and ${\beta}_{\pi K}$ in terms of
$r$,
\bea
{\beta}_{\pi\pi}&=& 1 + {2\over{r-1}}\,\bigg({{F_K^2}\over{F_{\pi}^2}} -
1\bigg) ( 1 + 2{\tilde\xi} /{\xi}) +\cdots\ ,\nonumber\\
{\beta}_{\pi K}&=& 1 + {2\over{r-1}}\,\bigg({{F_K^2}\over{F_{\pi}^2}} -
1\bigg)\big[ 1 + 2(3+r){\tilde\xi} /{\xi}\big]+\cdots\ ,\lbl{rbeta}
\ena
up to the ratio ${\tilde\xi}/{\xi}$, which is anyhow expected to be small, due
to the Zweig rule. Notice that $\xi$ and ${\tilde\xi}$ are related to the
low-energy constants $L_5$ and $L_4$ of Ref. \cite{GL}, respectively. Up to
higher order, $O(p^6 )$, corrections we may further
replace $r$ in the above formulae by its expression $r({\alpha}_{\pi\pi})$
obtained from Eq. \rf{lead}, but with ${\alpha}_{\pi\pi}^{lead}$ replaced
by ${\alpha}_{\pi\pi} = {\alpha}_{\pi\pi}^{lead} + {\delta}{\alpha}_{\pi\pi}$
(we take $F_K /F_{\pi}$ = 1.22).
Finally, we establish, in Sect. 3, the following sum-rule:
\eq
c =
- {5\over{144{\pi}^2}}\cdot{1\over{r-1}}\bigg\{
{\int}_{4M_{\pi}^2}^{\infty}\,{ds\over
s}\,[R_{+}(s) - 3R_{-}(s)]\,
-\,{1\over 2}\ln\bigg({{M_K}\over{M_{\pi}}}\bigg)\bigg\} + O(m_q^2,\, m_qB_0 )
\ ,\lbl{c}
\en
where $R_{+}$ ($R_{-}$) represents the cross section for $e^+ e^- $ into
hadronic final states with even (odd) total G-parity, normalized to the
cross section for $e^+ e^- \to {\mu}^+{\mu}^-$. Again, $r$ in the formula above
is to be thought as expressed in terms of ${\alpha}_{\pi\pi}$. The evaluation
of the integral in the narrow resonance approximation gives the following
estimate:
\eq
c = -\,{1\over{r-1}}(\, 4.6 \pm 2.3 )\times 10^{-3}\ .\lbl{cval}
\en
Hence, up to the small contribution from the kaon loops, and up to the Zweig
rule violations contained in $\zeta$ and in ${\tilde\xi}/\xi$, we have been
able to express the cross section for $\gamma\gamma\to\pi^0\pi^0$ at order
$O(p^5 )$ in terms of the single parameter ${\alpha}_{\pi\pi}$. This constant
could, in principle, be extracted from $\pi - \pi$ phase shifts, given
sufficiently accurate data\footnote{An analysis \cite{SSF} of the available
data shows that any value of ${\alpha}_{\pi\pi}$ in a range $\sim 1 - 4$ can be
accounted for.}. We recall that if standard $\chi$PT is correct, then
${\alpha}_{\pi\pi}$ is close to 1. The smaller $B_0$, the more
${\alpha}_{\pi\pi}$ increases towards ${\alpha}_{\pi\pi}\sim 4$. In Fig. 3,
we show the cross section for the three values
${\alpha}_{\pi\pi} = 1,\,3$ and 4. For ${\alpha}_{\pi\pi}\ge 3$, our result
agrees, within errors, with the data \cite{data} in the threshold region. For
${\alpha}_{\pi\pi} = 4$ ({\it i.e.} when the factor $1/(r-1)$ in \rf{rbeta} and
\rf{c} is largest), we have checked that the cross section is indeed
insensitive the the contribution coming from the kaon loops, by varying
${\alpha}_{\pi K}$ in the range $2\le {\alpha}_{\pi K} \le 6$, and by taking
${\beta}_{\pi K}$ from Eq. \rf{rbeta}. We have also checked that the result is
insensitive to the variation of the constant $c$ within the range $-1.30\times
10^{-3}\le c \le -0.44\times 10^{-3}$ given by Eq. \rf{cval} for
${\alpha}_{\pi\pi}=4$. Actually, even if
we set $c$ equal to zero, the low-energy cross section is only barely affected,
see Fig. 4. On the other hand, the cross section is more sensitive to
variations of ${\beta}_{\pi\pi}$ coming from different values of the Zweig rule
violating ratio ${\tilde\xi}/{\xi}$. Taking this ratio in the range $-0.2\le
{\tilde \xi}/{\xi} \le +0.2$, which amounts to a Zweig-rule violation of
40$\%$ in ${\beta}_{\pi\pi}$ (and corresponds to the range allowed for
$L_4 /L_5$ in \cite{GL}), we find
the variations in the cross section  as shown in Fig. 5.

\sect{A low-energy theorem for the $O(p^5 )$ constant $c$}

\indent

In this section, we derive the sum-rule \rf{c} and discuss its numerical
evaluation.
Consider the decomposition of the electromagnetic current into its $I$=1
and $I$=0 components:
\eq
j_{\mu} = V_{\mu}^3 + {1\over{\sqrt 3}}V_{\mu}^8\ ,\lbl{em}
\en
where
\eq
V_{\mu}^3 = {1\over 2}({\bar u}{\gamma}_{\mu}u-{\bar d}{\gamma}_{\mu}d)\ ,
\ V_{\mu}^8 = {1\over{2{\sqrt 3}}}({\bar u}{\gamma}_{\mu}u+{\bar
d}{\gamma}_{\mu}d
- 2{\bar s}{\gamma}_{\mu}s )\ .
\en
The corresponding vacuum polarization functions,
\eq
i\int {d^4}x\, e^{iq\cdot x}<{\Omega}\vert T\{ V_{\mu}^a (x) V_{\nu}^a
(0)\}\vert{\Omega}> = ( q_{\mu}q_{\nu} - {\eta}_{\mu\nu}q^2 ){\Pi}^{aa}(q^2 )\
\ ,\ a = 3,\,8\ ,\lbl{2pt}
\en
and
\eq
i\int {d^4}x\, e^{iq\cdot x}<{\Omega}\vert T\{ j_{\mu}^a (x) j_{\nu}^a
(0)\}\vert{\Omega}> = ( q_{\mu}q_{\nu} - {\eta}_{\mu\nu}q^2 ){\Pi}^{e.m.}(q^2 )
\ ,
\en
satisfy
\eq
{\Pi}^{e.m.}(Q^2 ) = {\Pi}^{33}(q^2 ) + {1\over 3}{\Pi}^{88}(q^2 )\ ,
\en
if isospin violations due to the electromagnetic interactions and to $m_u \ne
m_d$ are neglected.
Furthermore, the spectral density of the electromagnetic current is related to
the inclusive $e^+ e^-$ cross section
\eq
{1\over{\pi}}{\Im}m{\Pi}^{e.m}(q^2 ) = {1\over{12{\pi}^2}} R(q^2 )\ ,
\en
where, as usual,
\eq
R(q^2 ) = {{{\sigma}_{e^+ e^- \to had.}(q^2 )}\over{{\sigma}_{e^+ e^- \to
{\mu}^+{\mu}^-}(q^2 )}} = R_{+}(q^2 ) + R_{-}(q^2 )\ ,
\en
and where, corresponding to the decomposition \rf{em}, $R(q^2 )$ has been split
into its contributions from G-parity even and from G-parity odd hadronic final
states, with
\eq
{1\over{\pi}}{\Im}m{\Pi}^{33}(q^2 ) = {1\over{12{\pi}^2}}R_{+}(q^2 )\ \ ,\ \
{1\over{\pi}}{\Im}m{\Pi}^{88}(q^2 ) = {1\over{4{\pi}^2}}R_{-}(q^2 )\ .
\en
The difference ${\Pi}^{33}(q^2 ) - {\Pi}^{88}(q^2 )$ then satisfies an
unsubtracted dispersion relation, and consequently one can write the following
sum-rule:
\eq
{\Pi}^{33}(0 ) - {\Pi}^{88}(0 ) =
{1\over{{\pi}}}{\int}_{4M_{\pi}^2}^{\infty}\,{ds\over s}\,[{\Im}m{\Pi}^{33}(s)
- {\Im}m{\Pi}^{88}(s)]\ .\lbl{disprel}
\en
On the other hand, the left-hand side of Eq. \rf{disprel} above satisfies a
low-energy theorem: From the generating functional ${\tilde{\cal Z}}^{eff.}$ at
order $O(p^5 )$, one obtains:
\bea
{\Pi}^{33}(s)\,=\,&-&2[L_{10}^{r}({\mu}) + 2H_{1}^{r}({\mu})]\nonumber\\
                  &-&4(c_1 + c_2 + 2c_3 ){\hat m}\,-\,8c_4{\hat
m}(2+r)\nonumber\\
                  &+&{1\over 3s}(s-4M_{\pi}^2 ){\bar J}({s\over{M_{\pi}^2}})\,
+\,{1\over 6s}(s-4M_K^2 ){\bar J}({s\over{M_K^2}}) \lbl{pi33}\\
                 &-&{2\over 3}\cdot{1\over{32{\pi}^2}}
                   \bigg(\ln{{M_{\pi}^2}\over{{\mu}^2}} + 1 \bigg)\,
     -\, {1\over 3}\cdot{1\over{32{\pi}^2}}\bigg(\ln{{M_K^2}\over{{\mu}^2}}
+ 1\bigg)\,+\,{1\over{48{\pi}^2}}\,+\, O(s,\,{\hat m}^2,\,{\hat m}B_0 )
\ ,\nonumber\\
&{}&\nonumber\\
{\Pi}^{88}(s)\,=\,&-&2[L_{10}^{r}({\mu}) + 2H_{1}^{r}({\mu})]\nonumber\\
                  &-&{4\over 3}(1+2r)(c_1 + c_2 + 2c_3 ){\hat m}\,-\,8c_4{\hat
m}(2+r)\nonumber\\
                  &+&{1\over 2s}(s-4M_K^2 ){\bar J}({s\over{M_K^2}})
\lbl{pi88}\\
                  &-&{1\over{32{\pi}^2}}\bigg(\ln{{M_K^2}\over{{\mu}^2}}
+ 1\bigg)\,+\,{1\over{48{\pi}^2}}\,+\, O(s,\,{\hat m}^2,\,{\hat m}B_0 )
\ .\nonumber
\ena
The low-energy constants $L_{10}^{r}({\mu})$ and $H_{1}^{r}({\mu})$ are as
defined in \cite{GL}: they are common to both ${\cal L}^{(4)}$ and
${\tilde{\cal L}}^{(4)}$ since they are not associated with symmetry breaking
terms,
\eq
L_{10}\langle U^+F_{R\mu\nu}UF_{L}^{\mu\nu}\rangle + H_1\langle
F_{R\mu\nu}F_R^{\mu\nu} + F_{L\mu\nu}F_L^{\mu\nu}\rangle \in {\cal
L}^{(4)},\,{\tilde{\cal L}}^{(4)}\ .
\en
They depend on the renormalization scale $\mu$, again as specified in
\cite{GL}, but both ${\Pi}^{33}$ and ${\Pi}^{88}$ are independent of $\mu$. The
loop function ${\bar J}$ reads
\eq
 {\bar J}(z) \equiv
{1\over{16{\pi}^2}}\bigg\{\sigma\,\big(\,
\ln{{1-{\sigma}}\over{1+{\sigma}}} + i\pi\,\big)\, + 2\bigg\}\
 ,\ {\sigma}(z)\equiv{\sqrt{1-4/z}}\ ,
\en
for $z\ge 4$, and has the following expansion near $z=0$,
\eq
{\bar J}(z) =
{z\over{96{\pi}^2}} + O(z^2 )\ .
\en
Taking the difference of Eqs. \rf{pi33} and \rf{pi88}, and neglecting the
Zweig rule violating constant $c_4$ as in Eq. \rf{cst}, one obtains the
following sum-rule:
\bea
c&\sim& -{10\over 9}(c_1 + c_2 +2c_3 ){\hat m}\lbl{c2}\\
&=& - {5\over{12}}\cdot{1\over{r-1}}\,\bigg[\,{1\over{\pi}}\,
{\int}_{4M_{\pi}^2}^{\infty}\,{ds\over s}\,[{\Im}m{\Pi}^{33}(s) -
{\Im}m{\Pi}^{88}(s)]\,
- \,{1\over{24{\pi}^2}}\ln\bigg({{M_K}\over{M_{\pi}}}
\bigg)\,\bigg]\,+\,O({\hat m}^2,\,{\hat m}B_0 )\ ,\nonumber
\ena
which can now easily be written as given in Eq. \rf{c}. Notice also the
presence of the overall factor ${1/{(r-1)}}$ which suppresses the value
of $c$ when ${\alpha}_{\pi\pi}\sim 1$.
The above sum-rule is similar to and should be compared with the
sum-rule relating the scale invariant combination
\eq
{\bar L}_{10} \equiv L_{10}^{r}({\mu}) + {1\over{144{\pi}^2}} ( \ln M_{\pi}^2
/{\mu}^2\, +\, 1)
\en
to the difference of the spectral densities of the $I$ = 1 vector and
axial currents \cite{DH2}.

The right-hand side of Eq. \rf{c2} expresses $c$ as a difference of two
contributions. The first one, represented by the dispersive integral, is
numerically the most important one and positive, so that $c$ comes out with a
negative value. Writing this contribution as in Eq. \rf{c} suggests a direct
evaluation of the dispersive integral in terms of $e^+ e^-$ data. However,
such data are not so easy to obtain, especially in the isoscalar channel, and
are affected by large error bars. We refer to Ref. \cite{DG} for an overview
of the experimental situation and for references. For our part, we shall
evaluate the dispersive integral in the narrow resonance limit, taking into
account contributions from $\rho$(768), $\omega$(782) and $\phi$(1020):
\bea
{1\over{\pi}}\,
{\int}_{4M_{\pi}^2}^{\infty}\,{ds\over s}\,[{\Im}m{\Pi}^{33}(s) -
{\Im}m{\Pi}^{88}(s)]_{res.} &=&
{{3}\over{4\pi{\alpha}^2}}\,\bigg\{\, {{{\Gamma}_{\rho\to e^+
e^-}}\over{M_{\rho}}} - 3\cdot{{{\Gamma}_{\omega\to e^+ e^-}}\over{M_{\omega}}}
- 3\cdot{{{\Gamma}_{\phi\to e^+ e^-}}\over{M_{\phi}}}\,\bigg\}\nonumber\\
&\sim&\, (11.1\pm 2.0)\times 10^{-3}\ .\lbl{res}
\ena
The experimental values for the $e^+ e^-$ widths have been taken from the
Particle Data Group compilation \cite{PDG}, and the error comes from the
uncertainties on these numbers.
The second term on the right-hand side of Eq. \rf{c2} comes from contributions
of virtual 2$\pi$ and 2$K$ intermediate states to the vacuum
polarization functions ${\Pi}^{33}(0 )$ and ${\Pi}^{88}(0)$. Numerically,
we have
\eq
{1\over{24{\pi}^2}}\ln\bigg({{M_K}\over{M_{\pi}}}
\bigg) = 5.4\times 10^{-3}\ .
\en
We estimate the contributions of non-resonant states to the dispersive integral
by the size of this chiral logarithm. Indeed, if we evaluate the
dispersive integral \rf{res} replacing \hfill\break
${\Im}m{\Pi}^{33}(s) -
{\Im}m{\Pi}^{88}(s)$ by the imaginary parts as given by Eqs. \rf{pi33} and
\rf{pi88}, we precisely obtain this logarithm:
\eq
{1\over{\pi}}\,
{\int}_{4M_{\pi}^2}^{\infty}\,{ds\over s}\,[{\Im}m{\Pi}^{33}(s) -
{\Im}m{\Pi}^{88}(s)]_{\chi PT} =
{1\over{24{\pi}^2}}\ln\bigg({{M_K}\over{M_{\pi}}}\bigg)\ .
\en
Hence, we do not add it to
the contribution \rf{res}, but take it as our estimate of the uncertainty on
the
value of $c$, which we thus obtain as
\eq
c = -\,{1\over{r-1}}\cdot (4.6 \pm 2.3)\times 10^{-3}\ .
\en
For ${\alpha}_{\pi\pi} = 4$, one thus obtains $c = -(0.87\pm 0.43)\times
10^{-3}$, whereas for ${\alpha}_{\pi\pi} = 1$ this value comes out about five
times smaller, $c = -(0.18\pm 0.09)\times 10^{-3}$. This last value is relevant
for the comparision with the constant $d_3$ of Ref. \cite{BGS}. At first sight,
the fact that the latter is renormalized, whereas our $O(p^5 )$ constant is
finite, might appear as puzzling. The point is that the divergent part of $d_3$
comes with an additional power of $B_0$, and thus counts as order $O(p^6 )$ in
generalized $\chi$PT. Therefore, the relationship between these two constants
may be expressed as follows:
\eq
{d_3^r}(\mu ) = - {{9F_{\pi}^2 c}\over{160{\hat m}B_0}}\,\big\{ 1 +
O(B_0\ln\mu,\,
{\hat m}) + \cdots\,\big\}\ ,\lbl{d3}
\en
where the ellipses stand for additional corrections which might appear due to
the fact that the constant $c$ is defined in the three-light-flavours expansion
scheme. In order to compare with the value of $d_3^r$ extracted from standard
$\chi$PT, in the numerical evaluation of Eq. \rf{d3} above one should take the
value of $c$ for ${\alpha}_{\pi\pi}\approx 1$, corresponding to the standard
case, and
consistently replace, up to higher orders, 2${\hat m}B_0$ by $M_{\pi}^2$, {\it
i.e.} consider $B_0\sim O({\Lambda}_H )$. This gives (notice that the sign is
well defined) $d_3^r (\mu ) = (9.4\pm 4.7)\times 10^{-6} + O(B_0\ln\mu)+
\cdots$, which is not incompatible with, say, the value of the scale invariant
constant ${\bar d}_3$ as computed from Appendix D of Ref. \cite{BGS}:
\eq
{\bar d}_3 = {{F_{\pi}^2 C_S^{\gamma}C_S^m}\over{8M_S^2}} \sim \pm 3.9\times
10^{-6}\ .
\en
It is possible to discuss the relationships between, say, $\xi$ and $L_5$, or
between ${\tilde\xi}$ and $L_4$, along similar lines. This would, however, lead
us too far astray, and we defer this discussion to forthcoming publications
\cite{MS,KS}.

\indent

\sect{Comparison with the dispersive approach}

\indent

Morgan and Pennington \cite{MP} have devised a dispersion relation
treatment of the $\gamma\gamma\to\pi^0\pi^0$ amplitude which
proved capable of representing the data quite correctly. The
drawback of this type of approach is that the general
principles of unitarity and analyticity do not completely constrain
the amplitude. Donoghue and Holstein \cite{DH} argued that additional
constraints should be provided by the chiral expansion. In this section,
we discuss this idea, and we
use the dispersive method, in conjunction with our $O(p^5 )$
results, as a means of estimating the size of higher order chiral
corrections\footnote{We are indebted to J. Gasser for repeatedly emphasizing
this point to us.}.

In order to implement the method, one has to consider the charged
channel $\gamma\gamma\to\pi^+\pi^-$ as well as the neutral one. Let
us designate by $M^C_{\lambda\lambda'}$ and $M^N_{\lambda\lambda'}$
the corresponding helicity amplitudes, normalized as in Eq. (3), and
introduce the J=0 projections:
\eq
f_C(s)={1\over8\pi}\int d\Omega M^C_{++}\ , \qquad
f_N(s)={1\over8\pi}\int d\Omega M^N_{++}\ ,
\en
where $f_N(s)$ is related to the amplitude defined in
Eqs. (4), (5), $H(s,t,u)=H(s)+O(p^6)$, by  $H(s)=4f_N(s)/s$.
Next, one introduces isospin combinations $f_I(s),\ I=0,2$
\cite{MP} \cite{DH} which diagonalize the elastic unitarity relation:
\eq
f_C={2\over3}f_0+{1\over3}f_2\ , \qquad f_N={2\over3}(f_0-f_2)\ .
\en
Below the $K\bar K$ threshold, the following representation holds for
these functions \cite{MP}:
\eq
f_I(s)=l_I(s) +\Omega_I(s)s\bigg\{\gamma_I +(s_0-s)
{1\over\pi}\int^\infty_{4\mpi^2}\,{dx\over x}\,
{l_I(x){\Im}m(1/\Omega_I(x))\over (x-s_0)(x-s-i\epsilon)}\bigg\} \ ,
\lbl{disp}
\en
where $\Omega_I(s)$ is the Omn\`es-Mushkelishvili \cite{OM} function
constructed from the $\pi - \pi$ phase shifts ${\delta}_I (s)$ in the isospin
channel $I$,
\eq
\Omega_I(s)=\exp\left[{s\over\pi}\int_{4\mpi^2}^\infty\,{dx\over x}
{\delta_I(x)\over (x-s-i\epsilon)}\right]\ ,
\lbl{omega}
\en
(normalized such that $\Omega_I(0)=1$)
and $s_0$ is an arbitrary subtraction point.
The function $l_I(s)$ must have the same left-hand cut
structure as the amplitude $f_I(s)$ and be analytic in the
rest of the complex $s$ plane. The representation  \rf{disp}
then garantees that $f_I(s)$ has the correct analytic structure.
In particular, the
discontinuity along the right-hand cut $4\mpi^2<s<\infty$ is given by:
\eq
{\Im}m f_I(s)=f_I(s)e^{-i\delta_I(s)}\sin\delta_I(s)\ .
\en
In the left-hand cut function one can separate the single pion pole
(Born term)
\eq
l_I(s)={4\mpi^2\over s\sigma(s/\mpi^2)}\arctanh\,\sigma({s\over{\mpi^2}})
+s\tilde l_I(s)\ .
\lbl{lcut}
\en
The remaining piece, $\tilde l_I(s)$, is generated by two or more pion
exchanges in the t (u) channels and shows up in the chiral expansion
starting at order $O(p^6 )$.
A factor of $s$ has been separated in Eqs. \rf{disp} and  \rf{lcut}
on account
of low-energy electromagnetic theorems and a single arbitrary constant
(in each isospin channel) appears due to boundedness properties for
asymptotic values of $s$.

It should be clear that a rigorous evaluation of the  left-hand cut
function $\tilde l_I(s)$ is no simple matter.In fact, this can be done only
order by order in chiral perturbation theory.
For the purpose of doing
order of magnitude estimates, however,  it seems reasonable to
retain the contributions
of a few low-lying resonances: $\rho,\ \omega,\ a_1$
and $b_1$ \cite{KO}.
The low energy region of interest here, $2\mpi\le E_{\pi\pi}\le 450$
MeV proves to be fairly insensitive to the details of which resonances
are included and how they are parametrized \cite{MP}.
This leaves one with the task of determining the arbitrary constants
$\gamma_I$.  Following the suggestion of \cite{DH}, one should
be able to do so
by matching the dispersive representation  \rf{disp} with the \cpt
prediction, $f^\chi_I(s)$, in a point where the latter can be trusted. It is
of course
convenient to choose the subtraction point in  \rf{disp} equal to
the matching point, so that the constants $\gamma_I$ are simply
given by:
\eq
\gamma_I= {f^\chi_I(s_0)-l_I(s_0)\over s_0\Omega_I(s_0) }\ .
\lbl{match}
\en
One has
then to extend the \cpt calculation to the charged channel
and one finds, for the isospin functions $f^\chi_I(s)$:
\eq
f^\chi_I(s)={4\mpi^2\over s\sigma(s/\mpi^2)}\arctanh\,\sigma({s\over{\mpi^2}})+
A^\pi_I(s)\bar G({s\over\mpi^2})
+A^K_I(s)\bar G({s\over\mk^2})+{2(L_9+L_{10})s\over\fpi^2}-
{\Delta_I s\over\fpi^2}
\lbl{fi}
\en
where
\eq
A^\pi_0={1\over6\fpi^2}\big[\mpi^2(5\app-8\bpp)+6\bpp s\big]\ ,\qquad
A^\pi_2={1\over6\fpi^2}\big[\mpi^2(2\app+4\bpp)-3\bpp s\big]\ ,\
\en
\eq
A^K_0={1\over4\fpi^2}\bigg[
\bpik({3\over2}s-\mpi^2-\mk^2)+2\apik\mpi\mk +(\mk-\mpi)^2\bigg]
\ ,\qquad  A^K_2=0\ ,
\en
and where the function $\bar G$ was  defined in Eq. (11). $f_I^\chi$ contains
a contribution of order $O(p^2)+O(p^3)+O(p^4)$ as well as a
a set of contributions of chiral
order five which are collected in the constants  $\Delta_I$.
The combination $c=2(\Delta_0-\Delta_2)/3$ was evaluated in
Sec. 3.  In contrast,
we do not know of a simple way to estimate the combination
$c_+=(2\Delta_0+\Delta_1)/3$ which
contributes to the $\pi^+\pi^-$ channel. In principle, of course,
$\Delta_I$ can be evaluated from experiment via
\cpt analysis of the
reactions $\pi\to e\nu\gamma$,  $K\to e\nu\gamma$ and of the mean
charge radii of the pion and the of kaon.
One expects the size of $c_+$
to be of order 10-30\% that of $L_9+L_{10}$, so that it can be absorbed to
some extent into the uncertainty in the numerical value of
$L_9+L_{10}$ (note that for consistency this numerical value itself should be
determined
here from an $O(p^5)$ analysis).
We have included the contribution of the kaon loop
in  \rf{fi} for completeness, but since the corresponding $K\bar K$
discontinuity is not accounted for in the dispersive formula (this
could in principle be done following Ref. \cite{GM}),  we do not include
it when we perform the matching. It is worth noting that the chiral
formula  \rf{fi} satisfies the general dispersive representation:
at this order, one has to set $\tilde l_I(s)=0$, $\Omega_I(s)=1$ and
${\Im}m(1/\Omega_I(x))=-\sigma(s/\mpi^2)A^\pi_I(s)/16\pi$ in Eq. \rf{disp}.

One should be aware that there are some ambiguities in the dispersive
approach. Consider first the question of the matching point $s_0$
in  \rf{match}. The authors of Ref. \cite{DH}, for instance, have matched the
$O(p^4)$
\cpt amplitude with a dispersive representation not including the
vector meson contributions at $s=0$. A priori, one would expect that
any point in the region $0\le s_0\le 4\mpi^2$, where the amplitude is
real, should be as good as another. In practice though, the results are
sensitive to which point is chosen. We will show the results corresponding
to the two extreme cases $s_0=0$ and $s_0=4\mpi^2$.
A further source of uncertainty, in the dispersive approach, comes from
the $\pi - \pi$ phase-shifts themselves. Clearly, what will matter most
in the calculation is how the phase-shifts behave close to
threshold. This threshold behaviour is usually parametrized in terms of the
scattering
lengths and of the slope parameters.  Experimentally though, the phase-shifts
have been extracted only for energies larger than 600 MeV.
Extrapolation down to the threshold, even upon using constraints from
the Roy equations and from $Kl_4$ data, is subject to some uncertainty.
The influence of this uncertainty on the $\gamma\gamma\to\pi^0\pi^0$
dispersive amplitude has been considered in ref. \cite{MP}.
Furthermore, in our case, the matching requires that one knows the relation
between the constants $\app$, $\bpp$ and the set of phase-shifts. This
relationship has been studied in Ref. \cite{SSF} and we will rely on its
results in what follows.
Consider the  simple parametrization  proposed by  Schenk \cite{S}:
\eq
\tan\delta_I(s)={\sqrt{1-{{4\mpi^2}\over s}}}\,\bigg[
a_I +\tilde b_I ({s\over4\mpi^2}-1)+c_I({s\over4\mpi^2}-1)^2\,\bigg]
\left({4\mpi^2-E^2_I\over s-E^2_I}\right)\ ,
\en
where $\tilde b_I$ is related to the slope parameter $b_I$ by
$\tilde b_I=b_I-a_I 4\mpi^2/(E_I^2-4\mpi^2)+({a_I})^3$.
For $I=0$ we take two sets
of parameters resulting from two fits to the production data obtained by
fixing the value of the scattering length to $a_0=0.26$:
\bea
&&{\rm a})\ a_0=0.26,\ b_0=0.203,\ c_0=-0.0126,\ E_0=813.3 {\rm MeV}\lbl{a}\\
&&{\rm b})\ a_0=0.26,\ b_0=0.324,\ c_0=-0.0274,\ E_0=863.1 {\rm MeV}\ ,\lbl{b}
\ena
which correspond to the data of Eastabrooks-Martin \cite{EM}
and of Ochs \cite{OC}, respectively. The two sets differ
essentially by the value of the slope parameter $b_0$. The corresponding
values of our low-energy parameters are \cite{SSF}:
\eq
a)\ \app=3.35\qquad b)\ \app=4.10\ ,
\lbl{val}
\en
respectively, while $\bpp=1.17$ in both cases.
The isospin $I=2$ parameters are taken as in Ref. \cite{S}.
Given a matching point $s_0$, we now have all the ingredients needed
for a determination of the constants $\gamma_I$ via
 \rf{match} and the evaluation of the dispersive amplitudes from
 \rf{disp}.

In Figs. 6a and 6b we compare the dispersive evaluation of the amplitude
$H(s)$ (or more precisely $100\mpi^2H(s)$ for $0\le s\le4\mpi^2$ and
$100\mpi^2\vert H(s)\vert$ above threshold) for
$\gamma\gamma\to\pi^0\pi^0$, corresponding to the two parameter sets \rf{a} and
\rf{b} of $\pi - \pi$ phases to our $O(p^5)$ chiral calculation.
The experimental points correspond to the Crystal Ball data \cite{data}
under the assumption that only the S-wave contributes to the
cross section. One observes that in the case of the phases determined
from  \rf{a} (Fig. 6a)
the two dispersive representations (corresponding to
$s_0=0$ and $s_0=4\mpi^2$), the chiral calculation, as well as the
experimental data are all rather close to each other in all the
physical low-energy region of interest. On the contrary, in the case
of the parameters \rf{b} the dispersive amplitude always exceeds both
the experimental data and the perturbative amplitude. This is presumably
due to the fact that this set of phases is caracterized
by a slope parameter which is too large (the scattering length
being chosen to be $a_0=0.26$ in both cases). We note that the
$\gamma\gamma\to\pi^0\pi^0$ data provide only indirect information
on the low energy behaviour of the $\pi-\pi$ phases. On the other hand, the
chiral representation allows to constrain
the low energy parameters
$\app$ and $\bpp$, which are related in a non-trivial way to the
phases \cite{SSF}. One further observes that the difference between the
two dispersive curves with matching at $0$ and $4\mpi^2$, respectively,
is most significant around $s=0$. The amplitude is rather small in
this region and corrections appear to be large. For instance the
$O(p^6)$ contribution coming from the vector mesons is about 50\%
of the $O(p^5)$ chiral amplitude at $s=0$ , while it represents a
negligible 5\% at threshold.

Finally, we turn to the question of the pion polarizabilities. Due to our
lack of knowledge of the $O(p^5)$ corrections $\Delta_I$ (see  Eq. \rf{fi})
the charged pion polarizabilities can only be given to generalized
$O(p^4)$ order:
\eq
\bar\alpha_{\pi^+}={\alpha\over\mpi\fpi^2}\bigg[
{1\over144\pi^2}(\app-\bpp)+
{1\over576\pi^2}\lambda_K +4(L_9+L_{10})+O(p^5)\bigg]\ ,
\lbl{polc}
\en
where
\eq
\lambda_K=2(\apik-1){\mpi\over\mk}+(1-\bpik)(1+{\mpi^2\over\mk^2})\ .
\en
For the neutral pion, we find
\eq
\bar\alpha_{\pi^0}={\alpha\over\mpi\fpi^2}\,\bigg[
{1\over288\pi^2}(\app-4\bpp)+{1\over576\pi^2}\lambda_K
-2c\,\bigg]\ .
\lbl{poln}
\en
The difference with the standard O$(p^4)$ situation is that, firstly,
$\app$ and $\bpp$  (as well as $\apik$ and $\bpik$)
can depart significantly from the value 1  (the values used in the table
below are those quoted in  \rf{val} ).
Secondly, we have extra contributions which are of chiral order five.
In order to get a feeling of how significant
these differences are, we have collected a few numerical values
in Table (1) below:

\begin{center}
\begin{tabular}{|c|c|c||c|c|} \hline
                     & $\bar\alpha_{\pi^0}$&$\bar\beta_{\pi^0}$
                     & $\bar\alpha_{\pi^+}$&$\bar\beta_{\pi^+}$\\ \hline
standard $O(p^4)$ & -0.50 & 0.50 & 2.74 & -2.74 \\ \hline
standard $O(p^6)$\cite{BGS}        & -0.35 & 1.50 &      &       \\ \hline
generalized $O(p^5)$ &  0.44 &-0.44 & 3.47 & -3.47 \\ \hline
dispersive($s_0=4\mpi^2$)  & -0.76& 1.78 & 3.33 & -3.05 \\ \hline
dispersive($s_0=0$)  & 0.96 & 0.07 & 3.61 & -3.33  \\ \hline
\end{tabular}
\vskip 0.5 true cm
{\bf Table 1:} {\it Results for the electric $(\bar\alpha)$ and
magnetic $(\bar\beta)$ pion polarizabilities in \hfill\break
units of $10^{-4}$ fm$^3$.}
\end{center}
The dispersive results are evaluated using the phase-shift parameters
\rf{a}, and the values of $\app$, $\bpp$ used in the $O(p^5)$
expressions (63-65) correspond to the same phases (see (62a) ). We proceed as
follows. First, the difference in the electric and magnetic polarizations
are obtained from
\eq
\bar\alpha_{\pi^0} -\bar\beta_{\pi^0} ={\alpha\over\mpi}\lim_{s=0}
{4 f_N(s)\over s}\ ,
\en
($f_N$ being calculated from the representation \rf{disp}) and similarly
for the $\pi^+$ with $f_N$ replaced by $f_C$. Next, we assume that the
polarization sum is saturated by the spin one meson pole contributions (the
conventions and notations follow Ref. \cite{BGS}):
\eq
\bar\alpha_{\pi^0} +\bar\beta_{\pi^0} ={\alpha\over\mpi}\,8\mpi^2
\left({C_\rho\over M^2_\rho-\mpi^2} +
{C_\omega\over M^2_\omega-\mpi^2}+{C_{b_1}\over M^2_{b_1}-\mpi^2}\right)\ ,
\en
with $C_\omega=0.67,\ C_\rho=0.12,\ C_{b_1}=0.53$ (all in GeV$^{-2}$)
for the $\pi^0$, while for the charged pion the sum is given by the
same formula with $C_\omega=0,\ C_\rho=0.06,$ and $C_{b_1}$ as before.
These values are obtained from the experimental rates $V^0\to\pi^0\gamma$
and $V^+\to\pi^+\gamma$ respectively. Note the strong breaking of isospin
symmetry for the $\rho$ which is due to $\rho-\omega$ mixing.

The table shows that the difference between the standard and the
generalized \cpt is rather minor for the charged pion which, in both
cases, is dominated by $L_9+L_{10}$. Things are quite different for
the neutral pion since, in this case, we find the electric polarizability
to be positive. This result is essentially caused by the $O(p^5)$ constant
$c$ in formula \rf{poln}. This contribution is much larger than the
corresponding one in the $O(p^6)$ calculation of Ref. \cite{BGS} because
of the $1/(r-1)$ factor in (26) which is much larger in our case.
It is also striking that the dispersive
results with different matching points to the \cpt amplitude
yield rather different numbers for the polarizations.
Note finally that the kaon loop contribution, which
one would a priori expect to be negligible (and which is not
included in the numbers shown in the table) would further increase the
generalized $O(p^5 )$ value of ${\bar\alpha}_{\pi^0}$
by 20\%. This suggests that the neutral pion polarizabilities
may be as difficult to control theoretically as they are difficult to
measure experimentally.

\indent

\sect{Summary and concluding remarks}

\indent

{\bf i)} At the one loop order of generalized $\chi$PT, the
$\gamma\gamma\to\pi^0\pi^0$ amplitude consists of three parts which are all
separately finite: The one loop $O(p^4 )$ and $O(p^5 )$ parts, and the tree
level $O(p^5 )$ contribution, the latter representing a constant shift in the
amplitude $H(s)$. Neglecting kaon loops and Zweig rule violating effects
(described in standard $\chi$PT by the constants $L_4$ and $L_6$), the whole
amplitude can be expressed in terms of the parameter ${\alpha}_{\pi\pi}$ and of
the single constant $c$ describing the tree contribution of ${\tilde{\cal
L}}^{(5)}$. We have derived a low-energy theorem which relates the constant $c$
to the experimental cross section for $e^+e^-\to$hadrons and to
${\alpha}_{\pi\pi}$.

\indent

{\bf ii)} The constant ${\alpha}_{\pi\pi}$ is measurable in the low-energy $\pi
- \pi$ scattering \cite{SSF}. Its leading $O(p^2 )$ part
${\alpha}_{\pi\pi}^{lead}$ is a function of the quark mass ratio $r={m_s}/
{\hat m}$. Within the standard $\chi$PT, ${\alpha}_{\pi\pi}$ should remain
close to ${\alpha}_{\pi\pi}^{lead} = 1$ ($r\sim r_2 =25.9$). The generalized
$\chi$PT admits a substancially larger value of ${\alpha}_{\pi\pi}$, typically
${\alpha}_{\pi\pi}\approx
{\alpha}^{lead}_{\pi\pi}\lapprox
 4$, corresponding to
$r\gapprox r_1 =6.3$. For energies $E_{\pi\pi}<450$ MeV, the generalized
one loop cross section agrees with the Crystal Ball data \cite{data} within
errors, provided that ${\alpha}_{\pi\pi}\ge 3$ (implying $r<10$). The
low-energy cross section is barely affected by the uncertainties in the
determination of the constant $c$.

\indent

{\bf iii)} At order $O(p^5 )$ of generalized $\chi$PT, the neutral pion
polarizability $({\bar\alpha}-{\bar\beta})_{\pi_0}$ becomes positive, typically
$({\bar\alpha}-{\bar\beta})_{\pi_0} = (1.04\pm 0.60)\times 10^{-4}$fm$^3$ for
${\alpha}_{\pi\pi} =3$. This has to be compared with the values $-1\times
10^{-4}$fm$^3$ and $-1.85\times 10^{-4}$fm$^3$ predicted by the {\it standard}
$O(p^4 )$ and $O(p^6 )$ orders \cite{BGS}, respectively.
(${\bar\alpha}+{\bar\beta}=0$ up
to and including order $O(p^5 )$.) This result is due to the fact that in
generalized $\chi$PT the negative $O(p^4 )$ contribution is strongly suppressed
due to the proximity of the Adler zero, which for ${\alpha}_{\pi\pi}^{lead}\to
4$ moves towards the point $s = 0$. The polarizability is then dominated by the
positive $O(p^5 )$ tree contribution. On the other hand, our prediction for the
charged pion polarizabilities (Table 1) does not differ very much from the
standard $O(p^4 )$ result. Notice that in this case, we have no quantitative
control of the $O(p^5 )$ constant contribution.

\indent

{\bf iv)} To illustrate the convergence rate of the generalized $\chi$PT, one
may first compare the $O(p^4 )$ and $O(p^5 )$ contributions to the amplitude
$H(s)$. For ${\alpha}_{\pi\pi} = 3$, the value of $H(s)$ (multiplied by
$100M_{\pi}^2$) at threshold is 8.64, of which 7.23 come from $O(p^4 )$ loops,
0.81 from $O(p^5 )$ loops and the remaining 0.60 represent the $O(p^5 )$ tree
contribution. Next, in order to control whether, at sufficiently low energies,
the one loop generalized $\chi$PT correctly accounts for rescattering effects
and final state interactions, we have compared our $O(p^5 )$ perturbative
amplitude with a dispersion-theoretic amplitude which satisfies exact S-wave
unitarity. For ${\alpha}_{\pi\pi}\ge 3$ and for $\pi -\pi$ phase-shifts not
characterized by a too large slope parameter, we have found that the dispersive
and perturbative amplitudes do indeed agree reasonably well for
$E_{\pi\pi}<450$ MeV, and that there is no conflict with unitarity in this
energy range. These facts enforce the hope that higher orders of generalized
$\chi$PT will remain sufficiently small, such as not to spoil the agreement
with experiment.

\indent

It might be legitimate to ask that, at sufficiently low energies, $\chi$PT
should correctly reproduce experimental data without resorting to higher
orders, unless it is justified by kinematical reasons or by the high precision
of the data. In the case of the
reaction $\gamma\gamma\to\pi^0\pi^0$, the standard expansion of ${\cal
L}^{eff.}$ fails to satisfy this requirement. The present analysis suggests a
possible interpretation of this fact: At each order, the standard $\chi$PT
misses important {\it symmetry breaking} terms which are unduly relegated to
higher orders. In order to recover these terms and to reach agreement with
experiment, one then has to go up to unnaturally high orders of $\chi$PT. If
this conjecture turned out to be correct, similar phenomena could be expected
elsewhere (low-energy $\pi -\pi$ scattering, pion scalar form factors, symmetry
breaking aspects of $K_{l4}$ decays, etc.). In order to decide whether the
generalized $\chi$PT provides a relevant improvement of the standard expansion
of ${\cal L}^{eff.}$, and whether the values of the various low-energy
parameters ($B_0$, $r=m_s /{\hat m}$,...) should be revised correspondingly,
additional experimental informations are needed. More
precise data on low-energy $\gamma\gamma\to\pi^0\pi^0$ and on the related
process $\eta\to\pi^0\gamma\gamma$ will be welcome. Decisive information might
come from the model independent determination of low-energy $\pi -\pi$ phase
shifts and of $K_{l4}$ form factors at Da$\Phi$ne, as well as from a direct
measurement of light quark masses in exclusive tau decays \cite{SFK}.

\indent

\noindent{\bf Acknowledgements}

\indent

We thank J\"urg Gasser and Mikko Sainio for useful discussions.

\indent

\noindent{\Large{\bf Appendix}}

\indent

We have gathered, in this Appendix, a few formulae related to the discussion in
Section 2. The explicit expressions of the combinations $M_{\pi}^2
{\alpha}_{\pi\pi}$ and $(M_K -M_{\pi})^2 + 2M_{\pi}M_K {\alpha}_{\pi K}$ in
terms of the low-energy constants of ${\tilde{\cal L}}^{(2)}$ and of
${\tilde{\cal L}}^{(3)}$ read as follows:
\bea
M_{\pi}^2{\alpha}_{\pi\pi}&=&2{\hat m}B_0 + 4{\hat m}^2(4A_0 + 8Z_0^S +rZ_0^S )
\nonumber\\
&-& 4{\hat m}^2 B_0\big[ 3\xi + {\tilde\xi}(6 + r)\big]\nonumber\\
&-& 16{\hat m}^3 A_0\big[ 3\xi + 2{\tilde\xi}(3 + r)\big]
\nonumber\\
&-& 8{\hat m}^3 Z_0^S\big[ 3\xi (4+r) + {\tilde\xi}(24+14r+r^2)\big]\nonumber\\
&+& {\hat m}^3\big[ 81{\rho}_1 + {\rho}_2 + 2{\rho}_4 (82+16r+r^2 ) + {\rho}_5
(2+r^2 ) + 12{\rho}_7 (2+r)(14+r)\big]\nonumber\ ,
\ena
\newpage
and
\bea
(M_K - M_{\pi})^2 &+& 2M_{\pi}M_K{\alpha}_{\pi K}\,=\nonumber\\
&=& {\hat m}B_0 (3+r) + {\hat m}^2 A_0 (17+14r+r^2 ) + 2{\hat m}^2 Z_0^S
(18+17r+r^2 )\nonumber\\
&-& {\hat m}^2 B_0\big[ \xi (1+r)(11+r) + 2{\tilde\xi}(12+17r-r^2 )\big]
\nonumber\\
&-&{\hat m}^3 A_0\big[ \xi (1+r)(29+18r+r^2 ) +2{\tilde\xi}(40+71r+18r^2-r^3 )
\big]\nonumber\\
&-&{\hat m}^3 Z_0^S\big[ (1+r)(68+50r+2r^2 ) +2{\tilde\xi}(2+r)(64+42r-2r^2 )
\big]\nonumber\\
&+& 2{\hat m}^3\,\big[\ {3\over 4}{\rho}_1 (43+50r+14r^2 +r^3 )
\nonumber\\
& &\qquad + {1\over 4}{\rho}_2 (3+r^2 )\nonumber\\
& &\qquad + {1\over 2}{\rho}_4 (146+188r+59r^2 +3r^3 )\nonumber\\
& &\qquad + {1\over 4}{\rho}_5 (2+r^2)(3+r )\nonumber\\
& &\qquad + 3{\rho}_7 (2+r)(30+29r+r^2 )\,\big]\nonumber\ .
\ena
The order $O(p^3 )$ expressions for the pseudoscalar masses are given as:
\bea
M_{\pi}^2 &=& 2{\hat m}B_0 + 4{\hat m}^2 A_0 + 4{\hat m}^2 Z_0^S(2+r)
\nonumber\\
&-&4{\hat m}^2 B_0\big[ \xi + {\tilde\xi}(2+r)\big]\nonumber\\
&-&8{\hat m}^3 A_0\big[ \xi + {\tilde\xi}(2+r)\big]\nonumber\\
&-&8{\hat m}^3 (2+r)Z_0^S \big[ \xi + {\tilde\xi}(2+r)\big]\nonumber\\
&+& {\hat m}^3\,\big[\ 9{\rho}_1 + {\rho}_2\nonumber\\
& &\qquad + 2{\rho}_4 (10+4r+4r^2 ) + {\rho}_5 (2+r^2 )\nonumber\\
& &\qquad + 12{\rho}_7 (2+r)^2\,\big]\nonumber\ ,
\ena
and
\bea
M_K^2 &=& {\hat m}B_0 (1+r) + {\hat m}^2 A_0 (1+r)^2 + 2{\hat m}^2 Z_0^S
(2+r)(1+r)
\nonumber\\
&-&{\hat m}^2 (1+r)B_0\big[ \xi (1+r) + 2{\tilde\xi}(2+r)\big]\nonumber\\
&-&{\hat m}^3 (1+r)^2 A_0\big[ \xi (1+r) + 2{\tilde\xi}(2+r)\big]\nonumber\\
&-&2{\hat m}^3 (1+r)(2+r)Z_0^S\big[ \xi(1+r) + 2{\tilde\xi}(2+r)\big]
\nonumber\\
&+& {\hat m}^3 (1+r)\,\big[\ {3\over 2}{\rho}_1 (1+r+r^2 ) + {1\over 2}{\rho}_2
(1-r+r^2 )\nonumber\\
& &\qquad\qquad +3{\rho}_4(2+2r+r^2 ) + {1\over 2}{\rho}_5 (1+r+r^2 )
\nonumber\\
& &\qquad\qquad +6{\rho}_7 (2+r)^2\,\big]\nonumber\ .
\ena
Combining these formulae, one may obtain the expressions for
${\alpha}_{\pi\pi}$ and of ${\alpha}_{\pi K}$. At leading order, {\it i.e.}
when $\xi$, ${\tilde\xi}$, ${\rho}_1$,...${\rho}_7$ are set equal to zero, one
can express ${\hat m}B_0$ and ${\hat m}^2 A_0$ in terms of the quark mass ratio
$r$ and of the pseudoscalar masses:
\bea
{\hat m}B_0&=&{{M_{\pi}^2}\over{2(r^2 -1)}}\,\big[(r-r_1 )(r+r_1 +2) -
2{\zeta}(r_2 - r)(2+r)\big]\ ,\nonumber\\
{\hat m}^2 A_0 &=& {{M_{\pi}^2}\over{2(r^2 -1)}}\,(r_2 -r)\ ,\nonumber
\ena
from which one infers the expressions \rf{lead} for ${\alpha}_{\pi\pi}^{lead}$
and for ${\alpha}_{\pi K}^{lead}$.

\indent

\newpage

\indent

\noindent{\Large{\bf Figure captions}}

\indent

\noindent{\bf Figure 1}: The vertices from ${\tilde{\cal L}}^{(2)} +
{\tilde{\cal L}}^{(3)}$ and from ${\tilde{\cal L}}^{(5)}$ (for the last one)
which enter the computation of the order $O(p^5 )$ amplitude $H(s)$. In the
penultimate and antepenultimate graphs, the mass-shell conditions $p_1^2 =
p_2^2 = M_{\pi}^2$ for the outgoing ${\pi}^0$ lines were used. The constants
${\alpha}_{\pi\pi}$, ${\alpha}_{\pi K}$, ${\beta}_{\pi\pi}$, ${\beta}_{\pi K}$,
as well as the constant $c$ which appears in the last vertex, are defined in
the
text and in the Appendix.

\indent

\noindent{\bf Figure 2}: The cross section
${\sigma}(\gamma\gamma\to\pi^0\pi^0,\,\vert\cos\theta\vert\le Z)$ as a function
of the center of mass energy $E$ at order $O(p^4 )$ for $r = 25.9$ (dotted
line), $r = 10$ (dash-dotted line), and for $r = 6.3$ (solid line). The data
points are taken from Ref. \cite{data}, and the vertical error bars correspond
to the {\it statistical} errors alone.

\indent

\noindent{\bf Figure 3}: The cross section
${\sigma}(\gamma\gamma\to\pi^0\pi^0,\,\vert\cos\theta\vert\le Z)$ as a function
of the center of mass energy $E$ at order $O(p^5 )$ for ${\alpha}_{\pi\pi} = 1$
(dotted line), ${\alpha}_{\pi\pi} = 3$ (dash-dotted line) and for
${\alpha}_{\pi\pi} = 4$ (solid line), with ${\tilde\xi}/\xi = 0$ in all three
cases.

\indent

\noindent{\bf Figure 4}: The cross section
${\sigma}(\gamma\gamma\to\pi^0\pi^0,\,\vert\cos\theta\vert\le Z)$ as a function
of the center of mass energy $E$ for ${\alpha}_{\pi\pi} = 3$ (solid line) and
the same cross section, but with $c$ set equal to zero (dotted line).
In both cases, ${\tilde\xi}/\xi = 0$.

\indent

\noindent{\bf Figure 5}: The uncertainty in the order $O(p^5 )$ cross section
${\sigma}(\gamma\gamma\to\pi^0\pi^0,\,\vert\cos\theta\vert\le Z)$, for
${\alpha}_{\pi\pi} = 3$, coming from the variation of the Zweig rule violating
parameter ${\tilde\xi}/\xi$ between -0.2 (lower dotted curve) and +0.2 (upper
dotted curve). The solid curve corresponds to ${\tilde\xi}/\xi = 0$.

\indent

\noindent{\bf Figure 6}: The order $O(p^5 )$ amplitude $100 M_{\pi}^2 \vert
H(s)\vert$, for $s\ge 4M_{\pi}^2$, and$100 M_{\pi}^2 H(s)$, for
$0\le s\le 4M_{\pi}^2$), shown by the full line, as compared to the dispersive
results obtained by matching with the $\chi$PT expression at $s = 4M_{\pi}^2$
(dotted curve) or at $s = 0$ (dash-dotted curve) for the set of phase shifts
given by the parameters of Eq. \rf{a} (Fig. 6a) and of Eq. \rf{b} (Fig. 6b).

\newpage

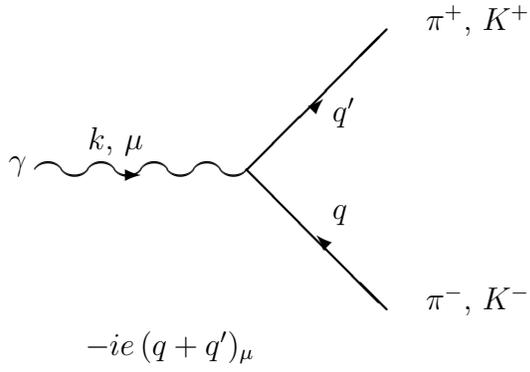
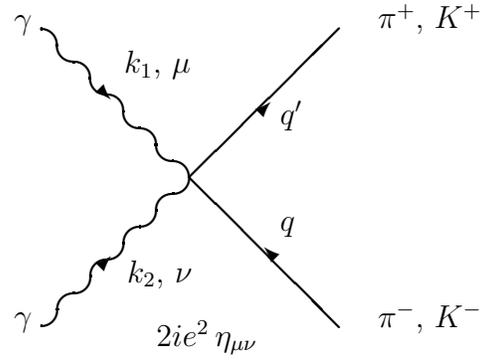
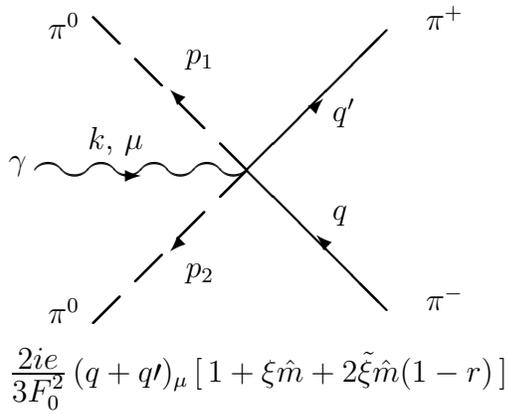
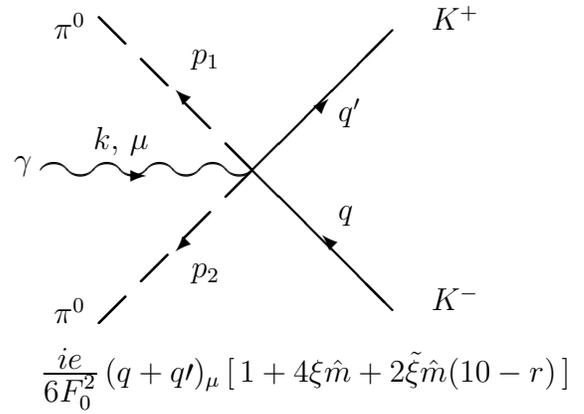
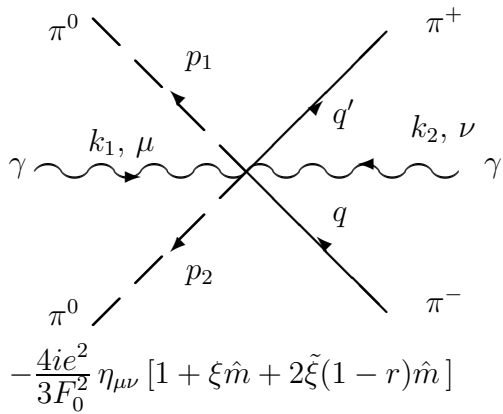
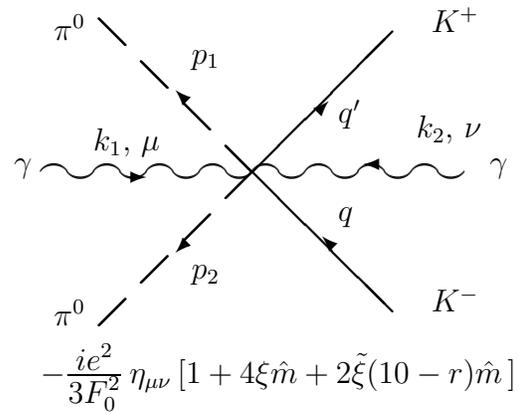
\input FEYNMAN

\indent

\indent

\vskip 2cm

\begin{picture}(100000,10000)
\bigphotons
\THICKLINES
\drawline\photon[\E\REG](1000,5000)[8]
\global\advance\pmidy by -200
\drawarrow[\E\ATTIP](\pmidx,\pmidy)
\put(0,5000){$\gamma$}
\global\advance\pmidx by -2000
\global\advance\pmidy by 1000
\put(\pmidx,\pmidy){$k$, $\mu$}
\drawline\fermion[\NE\REG](\photonbackx,\photonbacky)[7500]
\global\advance\pmidy by -200
\drawarrow[\LDIR\ATBASE](\pmidx,\pmidy)
\global\advance\pbackx by 300
\put(\pbackx,\pbacky){$\quad{\pi}^+$, $K^+$}
\global\advance\pmidx by 600
\global\advance\pmidy by -600
\put(\pmidx,\pmidy){$q^{\prime}$}
\drawline\fermion[\SE\REG](\photonbackx,\photonbacky)[7500]
\global\advance\pmidy by 200
\drawarrow[\NW\ATTIP](\pmidx,\pmidy)
\global\advance\pmidx by 600
\global\advance\pmidy by 600
\put(\pmidx,\pmidy){$q$}
\global\advance\pbackx by 300
\put(\pbackx,\pbacky){${\quad\pi}^-$, $K^-$}
\put(2500,-2000){ $-ie\,( q + q^{\prime}  )_{\mu}$}
\drawline\photon[\SE\REG](26000,10000)[9]
\global\advance\pmidx by -150
\global\advance\pmidy by 150
\drawarrow[\SE\ATTIP](\pmidx,\pmidy)
\put(25000,10000){$\gamma$}
\global\advance\pmidx by 500
\global\advance\pmidy by 1000
\put(\pmidx,\pmidy){$k_1$, $\mu$}
\drawline\fermion[\NE\REG](\photonbackx,\photonbacky)[8000]
\global\advance\pmidy by -200
\drawarrow[\LDIR\ATBASE](\pmidx,\pmidy)
\global\advance\pbackx by 300
\put(\pbackx,\pbacky){$\quad{\pi}^+$, $K^+$}
\global\advance\pmidx by 600
\global\advance\pmidy by -600
\put(\pmidx,\pmidy){$q^{\prime}$}
\drawline\fermion[\SE\REG](\photonbackx,\photonbacky)[8000]
\global\advance\pmidy by 200
\drawarrow[\NW\ATTIP](\pmidx,\pmidy)
\global\advance\pmidx by 600
\global\advance\pmidy by 600
\put(\pmidx,\pmidy){$q$}
\global\advance\pbackx by 300
\put(\pbackx,\pbacky){${\quad\pi}^-$, $K^-$}
\drawline\photon[\SW\REG](\photonbackx,\photonbacky)[9]
\global\advance\pmidx by -150
\global\advance\pmidy by -150
\drawarrow[\NE\ATTIP](\pmidx,\pmidy)
\put(25000,\pbacky){$\gamma$}
\global\advance\pmidx by 600
\global\advance\pmidy by -1000
\put(\pmidx,\pmidy){$k_2$, $\nu$}
\put(30000,-2000){ $2ie^2\,{\eta}_{\mu\nu}$}
\end{picture}

\vskip 4cm

\begin{picture}(40000,10000)
\bigphotons
\THICKLINES
\drawline\photon[\E\REG](1000,5000)[8]
\global\advance\pmidy by -200
\drawarrow[\E\ATTIP](\pmidx,\pmidy)
\put(0,5000){$\gamma$}
\global\advance\pmidx by -2000
\global\advance\pmidy by 1000
\put(\pmidx,\pmidy){$k$, $\mu$}
\drawline\fermion[\NE\REG](\photonbackx,\photonbacky)[7500]
\global\advance\pmidy by -200
\drawarrow[\LDIR\ATBASE](\pmidx,\pmidy)
\global\advance\pbackx by 300
\put(\pbackx,\pbacky){$\quad{\pi}^+$}
\global\advance\pmidx by 600
\global\advance\pmidy by -600
\put(\pmidx,\pmidy){$q^{\prime}$}
\drawline\fermion[\SE\REG](\photonbackx,\photonbacky)[7500]
\global\advance\pmidy by 200
\drawarrow[\NW\ATTIP](\pmidx,\pmidy)
\global\advance\pmidx by 600
\global\advance\pmidy by 600
\put(\pmidx,\pmidy){$q$}
\global\advance\pbackx by 300
\put(\pbackx,\pbacky){${\quad\pi}^-$}
\drawline\scalar[\NW\REG](\photonbackx,\photonbacky)[4]
\global\advance\pmidy by 150
\drawarrow[\NW\ATTIP](\pmidx,\pmidy)
\put(1500,10000){${\pi}^0$}
\global\advance\pmidx by 600
\global\advance\pmidy by 1000
\put(\pmidx,\pmidy){$p_1$}
\drawline\scalar[\SW\REG](\photonbackx,\photonbacky)[4]
\global\advance\pmidy by -150
\drawarrow[\SW\ATTIP](\pmidx,\pmidy)
\put(1500,\pbacky){${\pi}^0$}
\global\advance\pmidx by 600
\global\advance\pmidy by -1000
\put(\pmidx,\pmidy){$p_2$}
\put(0,-3000){$ {{\displaystyle 2ie}\over{\displaystyle 3F_0^2}}\,
                  ( q + q\prime  )_{\mu}\,
                 \big[\,1 + \xi{\hat m} + 2{\tilde\xi}{\hat m}(1 - r)\,\big]$}
\drawline\photon[\E\REG](26000,5000)[8]
\global\advance\pmidy by -200
\drawarrow[\E\ATTIP](\pmidx,\pmidy)
\put(25000,5000){$\gamma$}
\global\advance\pmidx by -2000
\global\advance\pmidy by 1000
\put(\pmidx,\pmidy){$k$, $\mu$}
\drawline\fermion[\NE\REG](\photonbackx,\photonbacky)[7500]
\global\advance\pmidy by -200
\drawarrow[\LDIR\ATBASE](\pmidx,\pmidy)
\global\advance\pbackx by 300
\put(\pbackx,\pbacky){$\quad K^+$}
\global\advance\pmidx by 600
\global\advance\pmidy by -600
\put(\pmidx,\pmidy){$q^{\prime}$}
\drawline\fermion[\SE\REG](\photonbackx,\photonbacky)[7500]
\global\advance\pmidy by 200
\drawarrow[\NW\ATTIP](\pmidx,\pmidy)
\global\advance\pmidx by 600
\global\advance\pmidy by 600
\put(\pmidx,\pmidy){$q$}
\global\advance\pbackx by 300
\put(\pbackx,\pbacky){$\quad K^-$}
\drawline\scalar[\NW\REG](\photonbackx,\photonbacky)[4]
\global\advance\pmidy by 150
\drawarrow[\NW\ATTIP](\pmidx,\pmidy)
\put(26500,10000){${\pi}^0$}
\global\advance\pmidx by 600
\global\advance\pmidy by 1000
\put(\pmidx,\pmidy){$p_1$}
\drawline\scalar[\SW\REG](\photonbackx,\photonbacky)[4]
\global\advance\pmidy by -150
\drawarrow[\SW\ATTIP](\pmidx,\pmidy)
\put(26500,\pbacky){${\pi}^0$}
\global\advance\pmidx by 600
\global\advance\pmidy by -1000
\put(\pmidx,\pmidy){$p_2$}
\put(26000,-3000){${{\displaystyle ie}\over{\displaystyle 6F_0^2}}\,
                ( q + q\prime  )_{\mu}\,
             \big[\,1 + 4\xi{\hat m} + 2{\tilde\xi}{\hat m}(10 - r)\,\big]$}
\end{picture}


\vskip 5cm

\begin{picture}(40000,10000)
\bigphotons
\THICKLINES
\drawline\photon[\E\REG](1000,5000)[8]
\global\advance\pmidy by -200
\drawarrow[\E\ATTIP](\pmidx,\pmidy)
\put(0,5000){$\gamma$}
\global\advance\pmidx by -2000
\global\advance\pmidy by 1000
\put(\pmidx,\pmidy){$k_1$, $\mu$}
\drawline\fermion[\NE\REG](\photonbackx,\photonbacky)[7500]
\global\advance\pmidy by -200
\drawarrow[\LDIR\ATBASE](\pmidx,\pmidy)
\global\advance\pbackx by 300
\put(\pbackx,\pbacky){$\quad{\pi}^+$}
\global\advance\pmidx by 600
\global\advance\pmidy by -600
\put(\pmidx,\pmidy){$q^{\prime}$}
\drawline\fermion[\SE\REG](\photonbackx,\photonbacky)[7500]
\global\advance\pmidy by 200
\drawarrow[\NW\ATTIP](\pmidx,\pmidy)
\global\advance\pmidx by 600
\global\advance\pmidy by 600
\put(\pmidx,\pmidy){$q$}
\global\advance\pbackx by 300
\put(\pbackx,\pbacky){${\quad\pi}^-$}
\drawline\scalar[\NW\REG](\photonbackx,\photonbacky)[4]
\global\advance\pmidy by 150
\drawarrow[\NW\ATTIP](\pmidx,\pmidy)
\put(1500,10000){${\pi}^0$}
\global\advance\pmidx by 600
\global\advance\pmidy by 1000
\put(\pmidx,\pmidy){$p_1$}
\drawline\scalar[\SW\REG](\photonbackx,\photonbacky)[4]
\global\advance\pmidy by -150
\drawarrow[\SW\ATTIP](\pmidx,\pmidy)
\put(1500,\pbacky){${\pi}^0$}
\global\advance\pmidx by 600
\global\advance\pmidy by -1000
\put(\pmidx,\pmidy){$p_2$}
\drawline\photon[\E\REG](\photonbackx,\photonbacky)[8]
\global\advance\pmidx by 200
\global\advance\pmidy by 300
\drawarrow[\W\ATTIP](\pmidx,\pmidy)
\put(18000,5000){$\gamma$}
\global\advance\pmidx by 2000
\global\advance\pmidy by 1000
\put(\pmidx,\pmidy){$k_2$, $\nu$}
\put(0,-3000){$-{{\displaystyle 4ie^2}\over{\displaystyle 3F_0^2}}\,
                {\eta}_{\mu\nu}
                \,\big[ 1 + \xi{\hat m} + 2{\tilde\xi}(1 - r){\hat m}\,\big]$}
\drawline\photon[\E\REG](26000,5000)[8]
\global\advance\pmidy by -200
\drawarrow[\E\ATTIP](\pmidx,\pmidy)
\put(25000,5000){$\gamma$}
\global\advance\pmidx by -2000
\global\advance\pmidy by 1000
\put(\pmidx,\pmidy){$k_1$, $\mu$}
\drawline\fermion[\NE\REG](\photonbackx,\photonbacky)[7500]
\global\advance\pmidy by -200
\drawarrow[\LDIR\ATBASE](\pmidx,\pmidy)
\global\advance\pbackx by 300
\put(\pbackx,\pbacky){$\quad K^+$}
\global\advance\pmidx by 600
\global\advance\pmidy by -600
\put(\pmidx,\pmidy){$q^{\prime}$}
\drawline\fermion[\SE\REG](\photonbackx,\photonbacky)[7500]
\global\advance\pmidy by 200
\drawarrow[\NW\ATTIP](\pmidx,\pmidy)
\global\advance\pmidx by 600
\global\advance\pmidy by 600
\put(\pmidx,\pmidy){$q$}
\global\advance\pbackx by 300
\put(\pbackx,\pbacky){$\quad K^-$}
\drawline\scalar[\NW\REG](\photonbackx,\photonbacky)[4]
\global\advance\pmidy by 200
\drawarrow[\NW\ATTIP](\pmidx,\pmidy)
\put(26500,10000){${\pi}^0$}
\global\advance\pmidx by 600
\global\advance\pmidy by 1000
\put(\pmidx,\pmidy){$p_1$}
\drawline\scalar[\SW\REG](\photonbackx,\photonbacky)[4]
\global\advance\pmidy by -150
\drawarrow[\SW\ATTIP](\pmidx,\pmidy)
\put(26500,\pbacky){${\pi}^0$}
\global\advance\pmidx by 600
\global\advance\pmidy by -1000
\put(\pmidx,\pmidy){$p_2$}
\drawline\photon[\E\REG](\photonbackx,\photonbacky)[8]
\global\advance\pmidx by 200
\global\advance\pmidy by 300
\drawarrow[\W\ATTIP](\pmidx,\pmidy)
\put(43000,5000){$\gamma$}
\global\advance\pmidx by 2000
\global\advance\pmidy by 1000
\put(\pmidx,\pmidy){$k_2$, $\nu$}
\put(26000,-3000){$-{{\displaystyle ie^2}\over{\displaystyle 3F_0^2}}
               \,{\eta}_{\mu\nu}
               \,\big[ 1 + 4\xi{\hat m} + 2{\tilde\xi}(10 - r){\hat m}\,\big]$}
\end{picture}

\vskip 4cm

\newpage

\indent

\indent

\vskip 2cm

\begin{picture}(40000,10000)
\THICKLINES
\drawline\scalar[\SE\REG](3000,10000)[4]
\global\advance\pmidy by 150
\drawarrow[\NW\ATTIP](\pmidx,\pmidy)
\put(1500,10000){${\pi}^0$}
\global\advance\pmidx by 600
\global\advance\pmidy by 1000
\put(\pmidx,\pmidy){$p_1$}
\drawline\fermion[\NE\REG](\scalarbackx,\scalarbacky)[7500]
\global\advance\pmidy by -200
\drawarrow[\LDIR\ATBASE](\pmidx,\pmidy)
\global\advance\pbackx by 300
\put(\pbackx,\pbacky){$\quad{\pi}^+$}
\global\advance\pmidx by 600
\global\advance\pmidy by -600
\put(\pmidx,\pmidy){$q^{\prime}$}
\drawline\fermion[\SE\REG](\scalarbackx,\scalarbacky)[7500]
\global\advance\pmidy by 200
\drawarrow[\NW\ATTIP](\pmidx,\pmidy)
\global\advance\pmidx by 600
\global\advance\pmidy by 600
\put(\pmidx,\pmidy){$q$}
\global\advance\pbackx by 300
\put(\pbackx,\pbacky){${\quad\pi}^-$}
\drawline\scalar[\SW\REG](\scalarbackx,\scalarbacky)[4]
\global\advance\pmidy by -150
\drawarrow[\SW\ATTIP](\pmidx,\pmidy)
\put(1500,\pbacky){${\pi}^0$}
\global\advance\pmidx by 600
\global\advance\pmidy by -1000
\put(\pmidx,\pmidy){$p_2$}
\put(3000,-4000){$i{{\displaystyle {\beta}_{\pi\pi}}\over{\displaystyle
F_{\pi}^2}}\,( s - {{\displaystyle 4}\over{\displaystyle 3}}M_{\pi}^2 ) +
i{{\displaystyle M_{\pi}^2{\alpha}_{\pi\pi}}\over{\displaystyle 3F_{\pi}^2}}$}
\put(-1000,-7500){$+{{\displaystyle i}\over{\displaystyle 3F_0^2}}(2M_{\pi}^2 -
q^2 - q^{{\prime}2})\big[ 1 + \xi{\hat m} + 2{\tilde\xi}{\hat m}(1-r)\big]$}
\drawline\scalar[\SE\REG](28000,10000)[4]
\global\advance\pmidy by 150
\drawarrow[\NW\ATTIP](\pmidx,\pmidy)
\put(26500,10000){${\pi}^0$}
\global\advance\pmidx by 600
\global\advance\pmidy by 1000
\put(\pmidx,\pmidy){$p_1$}
\drawline\fermion[\NE\REG](\scalarbackx,\scalarbacky)[7500]
\global\advance\pmidy by -200
\drawarrow[\LDIR\ATBASE](\pmidx,\pmidy)
\global\advance\pbackx by 300
\put(\pbackx,\pbacky){$\quad K^+$}
\global\advance\pmidx by 600
\global\advance\pmidy by -600
\put(\pmidx,\pmidy){$q^{\prime}$}
\drawline\fermion[\SE\REG](\scalarbackx,\scalarbacky)[7500]
\global\advance\pmidy by 200
\drawarrow[\NW\ATTIP](\pmidx,\pmidy)
\global\advance\pmidx by 600
\global\advance\pmidy by 600
\put(\pmidx,\pmidy){$q$}
\global\advance\pbackx by 300
\put(\pbackx,\pbacky){$\quad K^-$}
\drawline\scalar[\SW\REG](\scalarbackx,\scalarbacky)[4]
\global\advance\pmidy by -150
\drawarrow[\SW\ATTIP](\pmidx,\pmidy)
\put(26500,\pbacky){${\pi}^0$}
\global\advance\pmidx by 600
\global\advance\pmidy by -1000
\put(\pmidx,\pmidy){$p_2$}
\put(28000,-4000){$i{{\displaystyle {\beta}_{\pi K}}\over{\displaystyle
4F_{\pi}^2}}\,( s - {{\displaystyle 2}\over{\displaystyle 3}}M_{\pi}^2 -
{{\displaystyle 2}\over{\displaystyle 3}}M_K^2)$}
\put(26000,-7500){$ +{{\displaystyle i}\over{\displaystyle 6F_{\pi}^2}}\big[ (
M_K - M_{\pi})^2 + 2M_{\pi}M_K{\alpha}_{\pi K}\big]$}
\put(24500,-11000){$+{{\displaystyle i}\over{\displaystyle 12 F_0^2}}
(2M_K^2 - q^2 - q^{{\prime}2})\big[ 1 +
4\xi{\hat m} + 2{\tilde\xi}{\hat m}(10-r)\big]$}
\end{picture}

\vskip 6cm

\begin{picture}(40000,10000)
\bigphotons
\THICKLINES
\drawline\photon[\SE\REG](14000,10000)[9]
\global\advance\pmidx by -150
\global\advance\pmidy by 150
\drawarrow[\SE\ATTIP](\pmidx,\pmidy)
\put(13000,10000){$\gamma$}
\global\advance\pmidx by 600
\global\advance\pmidy by 1000
\put(\pmidx,\pmidy){$k_1$, $\mu$}
\drawline\scalar[\NE\REG](\photonbackx,\photonbacky)[4]
\global\advance\pmidy by -200
\drawarrow[\LDIR\ATBASE](\pmidx,\pmidy)
\global\advance\pbackx by 300
\put(\pbackx,\pbacky){$\quad\pi^0$}
\global\advance\pmidx by 600
\global\advance\pmidy by -600
\put(\pmidx,\pmidy){$p_1$}
\drawline\scalar[\SE\REG](\photonbackx,\photonbacky)[4]
\global\advance\pmidy by 200
\drawarrow[\LDIR\ATTIP](\pmidx,\pmidy)
\global\advance\pmidx by 600
\global\advance\pmidy by 600
\put(\pmidx,\pmidy){$p_2$}
\global\advance\pbackx by 300
\put(\pbackx,\pbacky){$\quad\pi^0$}
\drawline\photon[\SW\REG](\photonbackx,\photonbacky)[9]
\global\advance\pmidx by -150
\global\advance\pmidy by -150
\drawarrow[\NE\ATTIP](\pmidx,\pmidy)
\put(13000,\pbacky){$\gamma$}
\global\advance\pmidx by 600
\global\advance\pmidy by -1000
\put(\pmidx,\pmidy){$k_2$, $\nu$}
\put(12500,-4000){$-\,{{\displaystyle 4ie^2\,c}\over{\displaystyle
F_0^2}}\,\big[
k_1\cdot k_2{\eta}_{\mu\nu} - k_{1,\,\nu}k_{2,\,\mu}\big]$}
\end{picture}

\vskip 4cm

\begin{center}
{\Large Fig. 1}
\end{center}

\end{document}